\documentclass{article}

    \PassOptionsToPackage{numbers, compress}{natbib}

\usepackage[final]{neurips_2024}

\usepackage[utf8]{inputenc} 
\usepackage[T1]{fontenc}    
\usepackage{hyperref}       
\usepackage{url}            
\usepackage{booktabs}       
\usepackage{amsfonts}       
\usepackage{nicefrac}       
\usepackage{microtype}      
\usepackage{xcolor}         
\usepackage{placeins}       
\usepackage{microtype}
\usepackage{graphicx}
\usepackage{subfigure}
\usepackage{booktabs} 
\usepackage{pdflscape}
\usepackage{adjustbox}
\usepackage{float}
\usepackage{placeins}
\usepackage{longtable}
\usepackage{tablefootnote}

\usepackage{enumitem}
\usepackage{xspace}
\usepackage{xcolor}
\usepackage{wrapfig}
\usepackage{caption}
\usepackage{amssymb}
\usepackage{amsmath}
\usepackage{multirow}
\usepackage{adjustbox}
\usepackage{makecell}
\usepackage{subcaption}
\usepackage{algorithm}
\usepackage{algpseudocode}
\usepackage{colortbl}
\usepackage{mathtools}
\usepackage{setspace}
\usepackage{enumitem}
\usepackage{babel}
\usepackage{bbding}
\usepackage{utfsym}
\usepackage[flushleft]{threeparttable}

\usepackage{makecell}

\usepackage{thmtools}
\usepackage{thm-restate}

\definecolor{myred}{HTML}{e63946}
\definecolor{myblue}{HTML}{457b9d}

\setlist[enumerate]{leftmargin=7.5mm}

\usepackage{amsmath}
\usepackage{bm}

\def\eqref#1{equation~\ref{#1}}

\def\1{\bm{1}}

\DeclareMathAlphabet{\mathsfit}{\encodingdefault}{\sfdefault}{m}{sl}
\SetMathAlphabet{\mathsfit}{bold}{\encodingdefault}{\sfdefault}{bx}{n}

\let\oldAA\AA
\renewcommand{\AA}{\text{\normalfont\oldAA}}
\usepackage{cancel}

\title{Few-shot Protein Fitness Prediction via In-context Learning and Test-time Training}

\author{
\textbf{Felix Teufel}\textsuperscript{\rm 1,2,3}\;
\textbf{Aaron W. Kollasch}\textsuperscript{\rm 1}\;
\textbf{Yining Huang}\textsuperscript{\rm 1}\\
\textbf{Ole Winther}\textsuperscript{\rm 2,5}\;
\textbf{Kevin K. Yang}\textsuperscript{\rm 4}\;
\textbf{Pascal Notin}\textsuperscript{\rm 1,*}\;
\textbf{Debora S. Marks}\textsuperscript{\rm 1,*}\\
\textsuperscript{\rm *}{\small Corresponding Author} \\
\textsuperscript{\rm 1}{\small Department of Systems Biology, Harvard Medical School} \quad
\textsuperscript{\rm 2}{\small Department of Biology, University of Copenhagen} \\
\textsuperscript{\rm 3}{\small Machine Intelligence, Novo Nordisk A/S} \quad
\textsuperscript{\rm 4}{\small Microsoft Research, Cambridge, MA, USA} \\
\textsuperscript{\rm 5}{\small Dept. of Applied Mathematics and Computer Science, Technical University of Denmark} \\
\texttt{pascal\_notin@hms.harvard.edu}, \quad
\texttt{debbie@hms.harvard.edu} \\
}

\begin{document}

\maketitle

\begin{abstract}
Accurately predicting protein fitness with minimal experimental data is a persistent challenge in protein engineering. We introduce PRIMO (PRotein In-context Mutation Oracle), a transformer-based framework that leverages in-context learning and test-time training to adapt rapidly to new proteins and assays without large task-specific datasets. By encoding sequence information, auxiliary zero-shot predictions, and sparse experimental labels from many assays as a unified token set in a pre-training masked-language modeling paradigm, PRIMO learns to prioritize promising variants through a preference-based loss function. Across diverse protein families and properties—including both substitution and indel mutations—PRIMO outperforms zero-shot and fully supervised baselines. This work underscores the power of combining large-scale pre-training with efficient test-time adaptation to tackle challenging protein design tasks where data collection is expensive and label availability is limited.
\end{abstract}

\section{Introduction}
\label{sec:introduction}

Protein engineering has rapidly advanced in recent years, driven by breakthroughs in both experimental and computational methods. High-throughput (HT) experimental methods, such as Deep Mutational Scanning (DMS) assays~\citep{Fowler2014DeepMS}, enable large-scale exploration of sequence space by generating and testing thousands of variants for a desired function. 
The data generated through HT approaches can support the training of powerful machine learning (ML) models to learn the corresponding fitness landscape and further optimize the target properties~\citep{notin2023proteinnpt,groth2024kermut}. 
However, while HT assays have become more accessible for certain properties such as thermostability~\citep{Tsuboyama2023MegascaleEA,Beltran2024SitesaturationMO} or fluorescence~\citep{Somermeyer2021HeterogeneityOT}, it can still be prohibitively expensive, time-consuming, or altogether infeasible to produce large-scale functional measurements for many other properties.

Doing away with experimental annotations entirely, deep generative models such as Protein Language Models (pLMs) trained on large sets of natural sequences from protein data banks (eg., UniRef~\citep{Suzek2014UniRefCA}, MGnify~\citep{Richardson2022MGnifyTM}) have offered a promising avenue to address these shortcomings~\citep{meier2021language, notin2022tranception, nijkamp2023progen2}. However, although the zero-shot predictions they provide can be remarkably effective for protein design in certain settings~\citep{Russ2020AnEM}, they are still insufficiently accurate for many practical applications, providing rough starting points that need to be further optimized~\citep{Notin2024MachineLF}. 
As a result, \emph{few-shot} learning has emerged as a critical challenge in protein engineering: how can we accurately predict or optimize protein fitness from only a handful of experimental observations? 
Recent studies have attempted to tackle that problem by combining zero-shot scores with minimal labeled data \citep{biswas2021low, hawkins2024likelihood, groth2024kermut}. While this approach offers improved performance, such supervised methods often still demand a separate validation set to prevent overfitting, which can easily exceed the available budget in a strict few-shot setting (e.g., a single 96-well plate) and may fail to accommodate more complex variant types such as insertions and deletions.

To address these challenges, we introduce \textbf{PRIMO (PRotein In-context Mutation Oracle)}, a transformer-based framework that integrates \emph{in-context learning} with \emph{test-time training} to deliver highly accurate protein fitness predictions with only a handful of labeled samples per assay. PRIMO treats each sequence and any available measurements (including zero-shot predictions) as a unified token set in a masked language modeling paradigm, employing a preference-based loss to rank variants correctly. Pre-training over many assays allows PRIMO to adapt rapidly to new proteins and properties, circumventing the need for extensive labeled datasets or large validation sets. Moreover, PRIMO handles both substitution and indel mutations, broadening its applicability across a diverse range of protein engineering tasks.

Our main contributions are summarized as follows:
\begin{itemize}
\item \textbf{A new few-shot prediction framework}. We present PRIMO, a transformer architecture that combines in-context learning with test-time training, enabling accurate ranking of protein variants under extreme data scarcity.
\item \textbf{State-of-the-art few-shot performance}. We demonstrate that PRIMO significantly outperforms both zero-shot baselines and fully supervised models in low-data regimes, achieving superior fitness predictions even with a limited number of labeled examples, across diverse assays and protein families from the ProteinGym benchmark.
\item \textbf{Broad applicability}. Unlike many existing methods, PRIMO accommodates both single-substitution and indel variants and does not require a separate validation set, making it more practical for real-world protein design scenarios.
\item \textbf{A novel natural evolution benchmark}. We curate a benchmark comprised of several high-throughput assays that each characterize broad fitness landscapes spanned by natural sequences from a given protein family. This benchmark allows to assess models in challenging settings where train and test sequences are farther apart in sequence space.
\end{itemize}

Our results highlight the promise of large-scale pre-training on diverse deep mutational scans, followed by efficient test-time adaptation, in tackling challenging protein design tasks where experimental resources and labeled data are severely constrained.

\section{Related Work}
\label{sec:related}

As experimental budgets in biomolecule research can be highly constrained, few-shot learning for property prediction has been a long-standing challenge with high practical relevance.

\subsection{Zero-shot fitness prediction}

Given that learning from few-shot observations is challenging, a popular alternative approach is \textit{zero-shot} fitness prediction, where the likelihood of a model trained on evolutionarily observed sequences is used to score variant effects. Family specific models learn the distribution of evolutionary sequence within a protein family by leveraging Multiple Sequence Alignments (MSAs) \citep{hopf2017mutation, Frazer2021DiseaseVP}. Protein language models trained on large protein sequence databases capture the fitness distribution of many protein families in a single model without the need for MSAs \citep{meier2021language, notin2022tranception}. Augmenting sequence models with protein structure has been shown to further improve zero-shot prediction performance \citep{Zhang2024-s3f, Li2024-ProSST}.

Hybrid methods seek to unify the family-specific and protein language models. MSA Transformer \cite{Rao2021-msatransformer} learns from millions of MSAs across protein families. TranceptEVE \cite{Notin2022-TranceptEVE} complements the pre-trained protein language model with an alignment-based model at inference. To overcome the limitations of MSAs, PoET \cite{Truong2023-poet} proposed a retrieval-augmented model that scores sequences given a context set of related sequences, by concatenating and performing self-attention over the context sequences.

\subsection{Supervised learning of protein fitness}
 
 Shallow ML approaches such as ridge regression models on one-hot encoded amino acids or pLM embeddings have been successfully applied for few-shot protein engineering \citep{biswas2021low}, and see further performance boosts when incorporating zero-shot scores as additional features \citep{hsu2022learning}. Using DMS-scale datasets of protein fitness, ProteinNPT  \citep{notin2023proteinnpt} proposed to train transformer models for substitution variant effect prediction, also incorporating zero-shot scores as input features.

To further bridge the gap between unsupervised zero-shot prediction and  supervised learning from experimental labels, few-shot \textit{likelihood-based fine-tuning} of generative pLMs using a preference objective has been proposed by Hawkins-Hooker et al. \cite{hawkins2024likelihood}. 
 However, a key limitation of both fine-tuning and training ProteinNPT-scale models is the requirement of a validation set to prevent overfitting: Likelihood-based fine-tuning in \cite{hawkins2024likelihood} requires 128 observations for validation, which precludes true few-shot usage, and already exceeds the budget afforded by e.g. a single 96-well plate. 

Alternatively, rather than treating zero-shot scores as features, \textit{Kermut} \cite{groth2024kermut} proposed to use a zero-shot predictor as the prior mean function in a Gaussian process (GP) with a dedicated kernel for substitution variant effect prediction. As GPs are customarily trained using the likelihood of the training data, the architecture choice also circumvents the need for a validation dataset.

Machine learning guided Directed evolution \citep{Yang2025-DE} can also effectively explore protein fitness landscape with small number of wet-lab experiments. EVOLVEpro \cite{Jiang2025-EVOLVEpro} augments experimental directed evolution by combining Protein Language Models with few-shot learning. The efficient exploration of fitness landscape with machine learning could further be powered by automated robotic system for wet-lab experiments \cite{Rapp2024-SAMPLE}. 

Recently, Beck et al. \cite{beck2024metalic} proposed \textit{Metalic}, an in-context learning (ICL) approach using a model trained on many different proteins and fitness assays. As pure in-context learning proved too limiting to model new proteins, a supervised fine-tuning approach was adopted that again required 128 validation data points. Moreover, as Metalic reused the ProteinNPT transformer architecture, it cannot model indel variants. 
Lastly, Metalic was trained using a data split that is inadequate for transfer learning and yields high overlap between training and evaluation assays, as we will demonstrate in Section \ref{sec:bad_split}.

\section{The PRIMO Model}
\label{sec:method}

\begin{figure*}
    \centering
    \includegraphics[width=0.95\linewidth]{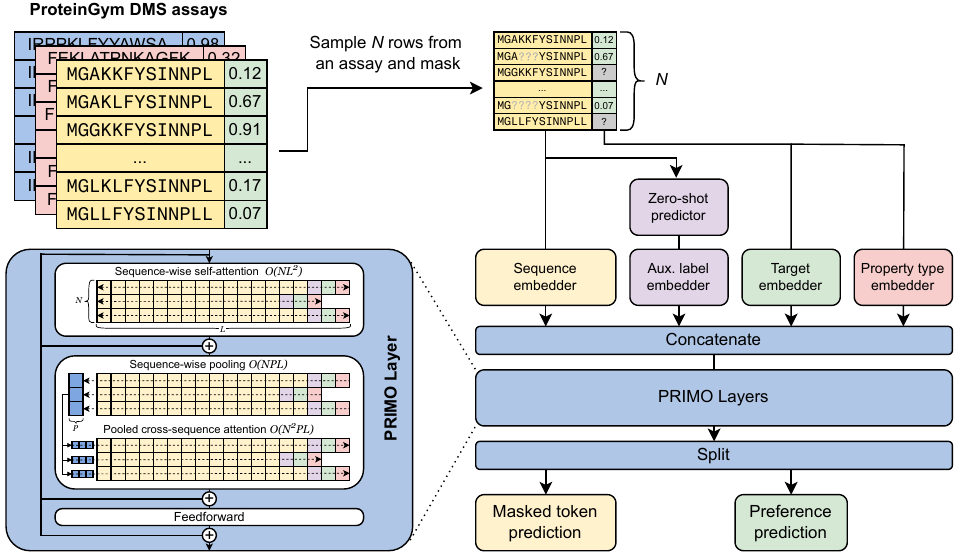}
    \caption{\textbf{The PRIMO architecture and training approach}. PRIMO processes labeled sets of proteins drawn from ProteinGym DMS assays. After processing the set with a transformer stack that allows for exchange of information between samples, it performs preference prediction on samples with masked fitness, and masked token prediction on amino acids.}
    \label{fig:pgformer}
\end{figure*}

\subsection{Architecture}

\subsubsection{Inputs}

PRIMO is a transformer-based masked language model that processes labeled sequence variant sets of size $N$. Each sample $i$ consists of an amino acid (AA) sequence $x^{\textup{AA}}_{i}$ of length $L$, a target quantitative fitness label $x^{\textup{Fitness}}_{i}$, a categorical property type ID $x^{\textup{ID}}_{i}$ denoting the assay type of the measured fitness, and one or multiple auxiliary zero-shot labels $x^{\textup{Auxiliary}}_{i}$. In PRIMO, we use autoregressive zero-shot scores from the ProGen pLM \citep{nijkamp2023progen2}, as they can be computed for both indel and substitution variants. 
All tokens are embedded and concatenated to a sequence $h_i$, using ESM-2 \citep{lin2023evolutionary} for the AA sequence and learned embeddings for all other inputs:

\begin{align}
    h^{\textup{AA}}_{i} &= \textup{ESM}(x^{\textup{AA}}_{i}) \nonumber\\
    h^{\textup{ID}}_{i} &= \textup{Embed}(x^{\textup{ID}}_{i}) \nonumber\\
    h^{\textup{Fitness}}_{i} &= \textup{Linear}(x^{\textup{Fitness}}_{i}) \nonumber\\
    h^{\textup{Auxiliary}}_{i} &= \textup{Linear}(x^{\textup{Auxiliary}}_{i}) \nonumber\\
    h_{i} &= \textup{Concat}(h^{\textup{AA}}_{i}, h^{\textup{Fitness}}_{i},  h^{\textup{ID}}_{i}, h^{\textup{Auxiliary}}_{i} ) 
\end{align}

\subsubsection{Transformer stack}

To overcome the computational complexity of full sequence-of-sequences self-attention, which suffers from quadratic scaling by the product of the sequence length and set size,
we employ a lightweight pooling attention mechanism to allow sequences to exchange information. 
Note that, as PRIMO processes sequences with indel mutations, column attention as used in ProteinNPT is less suitable, since insertions and deletions result in variable sequence lengths and misaligned positions. We first perform standard self-attention on each sequence $h_i$ individually, 
\begin{equation}
    {h_i} = \textup{MHA}(h_i,h_i,h_i) \quad \forall i \in {1, \ldots, N}.
\end{equation}

where $\textup{MHA}(q,k,v)$ denotes standard multi-head self-attention with query $q$, key $k$ and value $v$.
Next, we perform PRIMO's inter-sequence attention operation. We first pool each sequence $h_i$ into a  representation $p_i$ of fixed size $P$ using attention pooling \citep{dai2020funnel} (\autoref{eq:pool}). We concatetate the pooled representations of all sequences in the set (\autoref{eq:concat}), and let each sequence $h_i$ cross-attend to the pooled representations of all the sequences (\autoref{eq:mix}):



    

\begin{align}
    p_i &= \textup{MHA}(\textup{Mean}(h_i), h_i, h_i) \quad \forall i \in 1, \ldots, N\, \label{eq:pool}\\ 
    p &= \textup{Concat}(p_1, p_2, \ldots, p_N) \label{eq:concat} \\ 
    h_i &= \textup{MHA}(h_i, p,p ) \quad \forall i \in 1, \ldots, N. \label{eq:mix} \ 
\end{align}

As we exchange information between individual sequences only using pool representations, we can overcome the limiting scaling complexity of sequence-of-sequences self-attention, $\mathcal{O}(N^2L^2)$. The self-attention on each sequence has complexity of $\mathcal{O}(NL^2)$, while the sequence pooling and cross-attention has complexity of $\mathcal{O}(NPL)$ and $\mathcal{O}(N^2PL)$. As the size of the pooled representation used for cross-attention will typically be smaller than the sequence length, $NP << L$, the limiting attention scaling remains $\mathcal{O}(NL^2)$.
Together with a final feedfoward layer, the sequence-wise and cross-sequence attention operations constitute one PRIMO layer. 
As in PoET, we employ skip connections, pre-LayerNorm and rotary positional embeddings \citep{su2024roformer}. PRIMO uses 6 layers with a hidden size of 400.

\subsubsection{Prediction heads}

After the sequences have been processed by the PRIMO layers, we make predictions based on the updated hidden states $h^{\textup{AA}}_{i}$ and $h^{\textup{Fitness}}_{i}$. We reuse ESM's pre-trained prediction head, and train a linear layer for fitness prediction.

\begin{align}
    h^{\textup{AA}}_{i}, h^{\textup{Fitness}}_{i} &= \textup{Split}(h_{i}) \\
    \hat{x}^{\textup{AA}}_{i} &= \textup{ESMHead}(h^{\textup{AA}}_{i}) \\
    \hat{y}_{i} &= \textup{Linear}(h^{\textup{Fitness}}_{i})
\end{align}

\subsection{Pre-training}

PRIMO is pre-trained using a hybrid masked token reconstruction objective \citep{notin2023proteinnpt}: during pre-training, either the label $x^{\textup{Fitness}}_{i}$ or a span in $x^{\textup{AA}}_{i}$ may be masked. We mask labels with 33\% probability, and mask spans in the remaining samples with 20\% probability. As a set of sequences from a given DMS assay will be highly identical beyond a few mutations, making masked reconstruction mostly trivial, spans are placed such that the mutated regions are covered. 
For AA token reconstruction, we use a simple cross-entropy masked token prediction loss. For label reconstruction, following \cite{hawkins2024likelihood}, we use a preference-based loss that tasks the model with correctly ranking the fitness of the $Q$ masked samples in the set (non-masked samples are ignored from the loss calculation and serve as support context only):

\begin{equation}
    L = \sum_{i=1}^{Q} \sum_{j=1}^{Q} - \mathbb{I}(x^{\textup{Fitness}}_{i}>x^{\textup{Fitness}}_{j}) \log \sigma (\hat{y}_i - \hat{y}_j).
\end{equation}

$\mathbb{I}$ denotes the indicator function, and $\sigma$ is the sigmoid activation function. The loss is therefore equivalent to $Q \times Q$ binary classifications. For pre-training, we sample sets of size $N=32$ from one assay at a time, using a batch size of 12. Sequences are cropped to 512 AAs for computational efficiency. When cropping, we ensure that all relevant mutated positions are still present. The ESM-2 pLM for sequence embedding remains frozen during training.

\subsection{Test-time training}

After pre-training, we wish to predict the fitness of novel proteins given few-shot observations. As this usage may result in a distribution shift that makes in-context learning infeasible, we adopt a test-time training (TTT) \citep{pmlr-v119-sun20b, gandelsman2022testtime} approach for PRIMO. In TTT, instead of directly making predictions after conditioning on the context, the model's weights are first adapted to the task using fine-tuning on the context data. After predicting the test problem, the updated weights are discarded.
Rather than using a fully self-supervised objective for TTT, we take advantage of available few-shot observations and use PRIMO's hybrid sequence and preference label reconstruction loss, sampling masked sets from the few-shot data, as typically done for TTT in in-context learning scenarios \citep{akyürek2024surprisingeffectivenesstesttimetraining}. We perform gradient descent for a fixed number of 25 steps, using same loss function and learning rate as in pre-training. The pLM weights also remain frozen during TTT.

\subsection{Prediction}

For inference, we condition on $Q$ unmasked samples, and predict each test sample by masking it. The unnormalized preference score of each sample serves as its fitness prediction.
As PRIMO is a model that was designed to predict \textit{relative} fitness, we also need to provide context in zero-shot prediction that serves as the reference. We therefore provide an arbitrary sequence with an arbitrary fitness of 0.5 (center of a min-max scale) as unmasked context when scoring. Note that this is conceptually similar to zero-shot prediction with masked LMs, where a second sequence (usually the wild type) is required as reference.

We annotate all results obtained from predictions using direct conditioning on $Q$ samples as PRIMO (ICL), and any that first uses the same samples to perform TTT prior to conditioning as PRIMO (TTT).

\subsection{Training data}

We pre-train PRIMO on DMS assays from ProteinGym \citep{notin2024proteingym} that had a fitness readout that falls into the following categories: \textit{Stability, Enzymatic activity, Abundance, Fluorescence}, and \textit{Binding}. All assays that cover more specific aspects of function that do not fall into this categorization were excluded. We train on both substitution and indel variants. To overcome different experimental scales and units, raw DMS fitness values are min-max normalized for each assay. 


\begin{table*}[]
\caption{\textbf{Zero-shot prediction performance (Spearman correlation) of PRIMO when training on either the PRIMO or Metalic split}. The presence of training samples with high sequence identity to the test set in the Metalic split leads to inflated zero-shot performances.}
\vskip 0.15in
\label{tab:metalic}
\begin{center}
\begin{small}
    
\begin{tabular}{lllll}
\toprule
\multirow{2}{*}{Test dataset}                                                       & \multirow{2}{*}{\begin{tabular}[c]{@{}l@{}}Closest protein in \\ Metalic training set\end{tabular}} & \multirow{2}{*}{Identity} & \multicolumn{2}{l}{Zero-shot performance} \\
                                                                                    &                                                                                                       &                           & PRIMO split      & Metalic split      \\

\midrule
BLAT\_ECOLX\_Jacquier\_2013                                                         & BLAT\_ECOLX                                                                          & 100 \%    &    0.23                  &    0.65
                  \\
DYR\_ECOLI\_Thompson\_2019                                                          & DYR\_ECOLI                                                                           & 100 \%    &       0.45                &    0.50
                  \\
DLG4\_RAT\_McLaughlin\_2012                                                         & DLG4\_HUMAN                                                                          & 99 \%     &         
      0.30        &     0.37
                 \\
RL40A\_YEAST\_Roscoe\_2013                                                          & RL40A\_YEAST                                                                         & 100 \%    &           
       0.30     &               0.74
       \\
GFP\_AEQVI\_Sarkisyan\_2016                                                         & Q8WTC7\_9CNID                                                                        & 62 \%     &           0.20           &  0.44  \\
\bottomrule
\end{tabular}
\end{small}
\end{center}
\end{table*}

\section{Results}
\label{sec:results}

\subsection{Baselines}


Following Notin et al. \cite{notin2023proteinnpt}, we use zero-shot fitness augmented ridge regression models as the baseline. 
As demonstrated before \cite{hsu2022learning, biswas2021low}, such models can be used for few-shot learning without requiring a separate validation set to prevent overfitting during training. We also evaluate the EvolvePro \citep{Jiang2025-EVOLVEpro} random forest (RF) regression model that does not leverage zero-shot scores.
Additionally, we consider a GP with a zero-shot prior mean function and embedding kernel based on Kermut \citep{groth2024kermut}. We omit Kermut's structure kernel to enable modeling of indels.
To match the setup of PRIMO, all baseline models also use ESM-2 embeddings and ProGen zero-shot scores.


\subsection{Fitness prediction performance}
\label{sec:fewshot_results}

We draw an increasing number of $N$ labeled few-shot observations for learning, and report performance by predicting all samples in an assay and computing the Spearman rank correlation to the experimental fitness. We also perform zero-shot prediction at $N=0$, where we do not condition on any experimental measurement. We use a hold out subset of ProteinGym for evaluation that was designed to control for overlap to the training data on the protein level (\autoref{tab:hold_out}). 
We first trained two PRIMO models on this holdout ("PRIMO split") and the Metalic split that did not control for overlap.
In direct comparison, we find that the high sequence similarity overlap between Metalics' training and testing set can cause inflated zero-shot performances. 
In the most drastic case, on RL40A\_YEAST, where the Metalic pre-training set contains 2,633 observations of the same protein in two other assays, the 100\% similarity train-test overlap results in an apparent increase of 0.4 over the zero-shot performance obtained with the PRIMO split.

Using the PRIMO split, we first evaluate PRIMO's in-context learning capability. While zero-shot prediction shows an average improvement (0.51) over ProGen (0.41), performance mostly stays flat with increasing $N$ (\autoref{tab:main_results}). 
To rule out the possibility that PRIMO fails to process context,  effectively working as a single-sample model (a highly accurate single-sequence regression model would also perform well on a ranking metric), we ablate the inputs and only provide one unlabeled sequence at a time. This context-free prediction results in catastrophic failure, with performance becoming worse than random (\autoref{tab:reference_point}). This confirms that PRIMO did in fact learn to perform context-based prediction, and fundamentally works by \textit{comparing} sequences. We therefore consider it more likely that ICL is ineffective due to the pre-training data available in ProteinGym being too limiting to mitigate (expected) distribution shifts when testing on new assays. 

When using TTT to adapt PRIMO's parameters to the unseen test assays, we observe a gradual performance improvement with increasing data, going from a zero-shot average Spearman correlation of 0.51 to 0.67 with 128 shots. The GP and ridge regression baseline methods exhibit the same trend, but are outperformed by PRIMO with TTT on all levels of $N$. Especially at extreme low-$N$ of up to 32 shots, baseline methods prove ineffective, with the exception of the EvolvePro RF model on GFP fluorescence prediction. A detailed breakdown by substitution and indel mutations is provided in the appendix.



\begin{table*}[t]
\caption{\textbf{Few-shot prediction of held-out DMS assays}. Average Spearman correlation coefficient over all held out assays is shown, aggregated over five replicates. Per-assay performances are reported in \autoref{tab:results_1} and \autoref{tab:results_2}. The zero-shot performances of the GP and Ridge regression models marked with * are zero-shot predictions from ProGen that either serve as the prior mean (GP) or an additional input feature (Ridge). The best value per level of $N$ is highlighted in bold, unless zero-shot prediction is superior. Error bars are computed over different draws for each $N$.}
\label{tab:main_results}
\vskip 0.15in
\begin{center}
\begin{small}
{\fontsize{7.5pt}{9pt}\selectfont
\begin{tabular}{llllllll}
\toprule
Method                & \multicolumn{7}{l}{Shots}            \\
                      & 0  & 4  & 8 & 16   & 32   & 64   & 128  \\
\midrule
Overall\\
\midrule
GP & 0.42* & 0.24 ± 0.02 & 0.32 ± 0.03 & 0.39 ± 0.02 & 0.45 ± 0.01 & 0.51 ± 0.01 & 0.56 ± 0.00 \\
Ridge & 0.42* & 0.26 ± 0.02 & 0.33 ± 0.02 & 0.41 ± 0.01 & 0.48 ± 0.01 & 0.56 ± 0.01 & 0.63 ± 0.00 \\
RF & - & 0.23 ± 0.01 & 0.32 ± 0.01 & 0.39 ± 0.01 & 0.45 ± 0.02 & 0.52 ± 0.00 & 0.59 ± 0.00 \\
PRIMO (ICL) & \textbf{0.51 ± 0.01} & \textbf{0.53 ± 0.00} & \textbf{0.53 ± 0.00} & 0.53 ± 0.00 & 0.53 ± 0.00 & 0.53 ± 0.00 & 0.53 ± 0.00 \\
PRIMO (TTT) & \textbf{0.51 ± 0.01} & 0.49 ± 0.01 & 0.51 ± 0.01 & \textbf{0.54 ± 0.01} & \textbf{0.58 ± 0.02} & \textbf{0.63 ± 0.00} & \textbf{0.67 ± 0.01} \\

\midrule
Stability\\
\midrule
GP & 0.41 & 0.36 ± 0.05 & 0.47 ± 0.04 & 0.55 ± 0.02 & 0.61 ± 0.01 & 0.66 ± 0.01 & 0.71 ± 0.00 \\
Ridge & 0.41 & 0.39 ± 0.03 & 0.46 ± 0.04 & 0.55 ± 0.02 & 0.61 ± 0.01 & 0.69 ± 0.01 & 0.75 ± 0.00 \\
RF & - & 0.36 ± 0.03 & 0.46 ± 0.02 & 0.53 ± 0.01 & 0.58 ± 0.02 & 0.65 ± 0.01 & 0.71 ± 0.00 \\
PRIMO (ICL) & \textbf{0.61 ± 0.01} & \textbf{0.62 ± 0.00} & \textbf{0.62 ± 0.00} & 0.62 ± 0.00 & 0.62 ± 0.00 & 0.62 ± 0.00 & 0.62 ± 0.00 \\
PRIMO (TTT) & 0.59 ± 0.01 & 0.59 ± 0.02 & 0.62 ± 0.02 & \textbf{0.65 ± 0.01} & \textbf{0.69 ± 0.01} & \textbf{0.73 ± 0.01} & \textbf{0.77 ± 0.01} \\
\midrule
Enzymatic activity \\
\midrule
GP & 0.52* & 0.11 ± 0.06 & 0.19 ± 0.04 & 0.25 ± 0.03 & 0.31 ± 0.01 & 0.37 ± 0.02 & 0.42 ± 0.01 \\
Ridge & 0.52* & 0.11 ± 0.05 & 0.21 ± 0.03 & 0.26 ± 0.02 & 0.35 ± 0.03 & 0.44 ± 0.02 & 0.51 ± 0.01 \\
RF & - & 0.07 ± 0.05 & 0.17 ± 0.02 & 0.23 ± 0.04 & 0.3 ± 0.02 & 0.38 ± 0.02 & 0.47 ± 0.01 \\
PRIMO (ICL) & 0.49 ± 0.05 & \textbf{0.57 ± 0.00} & \textbf{0.57 ± 0.00} & \textbf{0.57 ± 0.00} & \textbf{0.57 ± 0.00} & \textbf{0.57 ± 0.00} & 0.57 ± 0.00 \\
PRIMO (TTT) & \textbf{0.53 ± 0.02} & 0.44 ± 0.05 & 0.49 ± 0.01 & 0.51 ± 0.01 & 0.54 ± 0.03 & 0.56 ± 0.01 & \textbf{0.61 ± 0.01} \\

\midrule
Fluorescence\\
\midrule
GP & 0.09* & 0.01 ± 0.08 & 0.02 ± 0.09 & 0.09 ± 0.01 & 0.14 ± 0.01 & 0.17 ± 0.02 & 0.24 ± 0.01 \\
Ridge & 0.09* & 0.08 ± 0.05 & 0.08 ± 0.07 & 0.12 ± 0.05 & 0.16 ± 0.02 & 0.21 ± 0.02 & 0.28 ± 0.02 \\
RF & - & 0.10 ± 0.08 & \textbf{0.12 ± 0.04} & \textbf{0.15 ± 0.03} & \textbf{0.21 ± 0.03} & 0.24 ± 0.01 & \textbf{0.32 ± 0.01} \\
PRIMO (ICL) & 0.07 ± 0.00 & 0.06 ± 0.00 & 0.06 ± 0.00 & 0.06 ± 0.00 & 0.06 ± 0.00 & 0.06 ± 0.00& 0.06 ± 0.00 \\
PRIMO (TTT) & \textbf{0.11 ± 0.02} & \textbf{0.14 ± 0.06} & 0.11 ± 0.02 & \textbf{0.15 ± 0.05} & 0.20 ± 0.04 & \textbf{0.25 ± 0.02} & 0.30 ± 0.02 \\
\midrule
Binding\\
\midrule
GP & 0.47* & 0.18 ± 0.04 & 0.22 ± 0.05 & 0.28 ± 0.03 & 0.34 ± 0.05 & 0.41 ± 0.02 & 0.46 ± 0.01 \\
Ridge & 0.47* & 0.17 ± 0.05 & 0.24 ± 0.04 & 0.33 ± 0.04 & 0.42 ± 0.03 & 0.51 ± 0.01 & 0.58 ± 0.01 \\
RF & - & 0.15 ± 0.05 & 0.22 ± 0.02 & 0.3 ± 0.03 & 0.38 ± 0.05 & 0.46 ± 0.01 & 0.55 ± 0.01 \\
PRIMO (ICL) & \textbf{0.49 ± 0.00} & \textbf{0.50 ± 0.00} & \textbf{0.50 ± 0.00} & \textbf{0.50 ± 0.00} & 0.50 ± 0.00 & 0.50 ± 0.00 & 0.50 ± 0.00 \\
PRIMO (TTT) & 0.47 ± 0.01 & 0.42 ± 0.05 & 0.42 ± 0.04 & 0.48 ± 0.02 & \textbf{0.56 ± 0.04} & \textbf{0.63 ± 0.02} & \textbf{0.69 ± 0.01} \\

\bottomrule
\end{tabular}
}
\end{small}
\end{center}
\end{table*}

\subsection{Performance on the natural evolution benchmark}

The previous analysis focused on test assays from ProteinGym in which mutated sequences vary from the reference wild-type sequence by at most a handful of mutations (typically singles or doubles). In order to assess the ability of models to extrapolate farther away in sequence space, we curate a new benchmark, comprised of three high-throughput assays that each characterize broad fitness landscapes spanned by natural sequences for Chorismate mutase~\citep{Russ2020AnEM}, Rubisco~\citep{Davidi2020HighlyAR} and PPAT~\citep{Plesa2018} respectively. We find that PRIMO outperforms all other baselines (Table~\ref{tab:evo_landscapes}). However, in that context, it is critical to make use of test-time training to allow the model to adapt more flexibly to the broader landscapes characterized by the test set.

\begin{table}[ht]
\centering
\caption{\textbf{Performance (Spearman correlation) on the natural evolution benchmark}. Values are reported as mean and standard deviation over five replicates for three assays. The zero-shot performances of the GP and Ridge regression models marked with * are zero-shot predictions from ProGen that either serve as the prior mean (GP) or an additional input feature (Ridge).}
\vskip 0.15in
\label{tab:evo_landscapes}
\begin{tabular}{lccccc}
\toprule
Shots & 0 & 4 & 8 & 16 & 32 \\
\midrule
PRIMO ICL & \textbf{0.10 $\pm$ 0.01} & \textbf{0.09 $\pm$ 0.02} & 0.09 $\pm$ 0.02 & 0.07 $\pm$ 0.03 & 0.06 $\pm$ 0.01 \\
PRIMO TTT & 0.09 $\pm$ 0.01 & 0.07 $\pm$ 0.03 & \textbf{0.10} $\pm$ 0.03 & \textbf{0.19 $\pm$ 0.06} & \textbf{0.30 $\pm$ 0.02} \\
GP & 0.04* & 0.02 $\pm$ 0.04 & 0.04 $\pm$ 0.04 & 0.07 $\pm$ 0.02 & 0.18 $\pm$ 0.11 \\
Ridge & 0.04* & 0.00 $\pm$ 0.03 & 0.08 $\pm$ 0.04 & 0.14 $\pm$ 0.02 & 0.24 $\pm$ 0.12 \\
RF & -- & 0.03 $\pm$ 0.03 & 0.08 $\pm$ 0.06 & 0.13 $\pm$ 0.06 & 0.24 $\pm$ 0.07 \\
MLP & -- & 0.00 $\pm$ 0.03 & 0.07 $\pm$ 0.04 & 0.13 $\pm$ 0.01 & 0.24 $\pm$ 0.11 \\
\bottomrule
\end{tabular}
\end{table}

\section{Discussion}
\label{sec:discussion}

\paragraph{Inappropriate splitting inflates prediction performance.} \label{sec:bad_split}

Data partitioning strategies are critical to ensure reliable performance reporting in ML on biological sequences, as widely recognized in the field \citep{teufel2023graphpart,bernett2024guiding,ektefaie2024evaluating}. When exploring a new paradigm such as pre-training on DMS data, followed by ICL evaluation, previous ProteinGym test subsets are inadequate, as they were only designed for single-assay learning and evaluation. As one may expect from finding cases of 100\% sequence identity overlap between partitions, we observed that apparent "zero-shot" performances can be highly inflated, as they in fact are driven by thousands of training observations of the same property of the same protein, just measured in a different experiment.
While Metalic did not report per-assay performances, we believe that its claimed average performance of 0.484 suffers from the issue demonstrated in \autoref{tab:metalic}, and find that Metalic underperforms PRIMO when training on a clean split (\autoref{tab:metalic_vs_primo}).

\paragraph{TTT can boost prediction performance.}
In our experiments, we find that ICL often fails to efficiently use the provided context, with performance staying flat and there not being a marginal benefit from providing more labeled data. While it remains to some degree unclear why ICL is not effective, it needs to be recognized that using ProteinGym as the pre-training dataset is limiting, only exposing PRIMO to context sets sampled from 116 distinct experiments. However, we find that TTT  can be an effective remedy to the unavoidable distribution shift at test time, allowing the model to adapt to new tasks in weight space as more data becomes available. 





\section{Conclusion and Outlook}
\label{sec:conclusion}

In this work, we have introduced PRIMO, a transformer model for in-context learning with test-time training that enables few-shot protein fitness prediction. By training on a large number of protein fitness assays, PRIMO learns to extract information from labeled samples and perform relative protein fitness prediction. 
While the direct application of ICL can be limiting and fail to efficiently use the available labeled information, TTT enables PRIMO to adapt to unseen assays, making it an efficient few-shot learner that achieves state of the art performance. 
PRIMO demonstrates that while the problem remains challenging, progress can be made by leveraging existing experimental data.

From a modeling perspective, assuming such a high-diversity, low-$N$ data resource can be established, a more flexible encoding of property types could be applied to widen the scope. Given sufficient metadata of the actual experimental protocols, natural language encoding could be considered. 
Moreover, future work may find it useful to consider leveraging other pLMs than ESM-2 in PRIMO, or selectively fine-tune the pLM, as it has been demonstrated that doing so can aid prediction performance in some cases \citep{gordon2025protein}.

In summary, our work demonstrates the potential of the ICL paradigm for few-shot fitness prediction, which we believe to become increasingly relevant in the future, given the success of ICL and large-scale pre-training in general across domains \citep{dong2024surveyincontextlearning, hollmann2025accurate}. 
We highlight the need for dedicated  fit-for-purpose data splitting regimes and detailed performance reporting, which we consider to be crucial for further advancement of the field.





\section{Availability}

Code and data is available at \url{https://github.com/fteufel/PRIMO}.

\section*{Acknowledgements}

We thank Peter M\o rch Groth for helpful discussions regarding Kermut. 


\bibliography{reference}

@INPROCEEDINGS{Rao2021-msatransformer,
  title     = "{MSA} Transformer",
  author    = "Rao, Roshan M and Liu, Jason and Verkuil, Robert and Meier,
               Joshua and Canny, John and Abbeel, Pieter and Sercu, Tom and
               Rives, Alexander",
  booktitle = "International Conference on Machine Learning",
  publisher = "PMLR",
  pages     = "8844--8856",
  month     =  jul,
  year      =  2021,
  language  = "en"
}

@article{Zhang2024-s3f,
  title     = "Multi-Scale Representation Learning for Protein Fitness
               Prediction",
  author    = "Zhang, Zuobai and Notin, Pascal and Huang, Yining and Lozano,
               Aurelie and Vijil, Vijil and Marks, Debora and Das, Payel and
               Tang, Jian",
  journal = "Annual Conference on Neural Information Processing Systems",
  year      =  2024,
  language  = "en"
}

@ARTICLE{Li2024-ProSST,
  title    = "{ProSST}: Protein language modeling with quantized structure and
              disentangled attention",
  author   = "Li, Mingchen and Tan, Pan and Ma, Xinzhu and Zhong, Bozitao and
              Yu, Huiqun and Zhou, Ziyi and Ouyang, Wanli and Zhou, Bingxin and
              Hong, Liang and Tan, Yang",
  journal  = "bioRxiv",
  pages    = "2024.04.15.589672",
  month    =  apr,
  year     =  2024
}

@ARTICLE{Truong2023-poet,
  title         = "{PoET}: A generative model of protein families as
                   sequences-of-sequences",
  author        = "Truong, Jr, Timothy F and Bepler, Tristan",
  journal       = "37th Conference on Neural Information Processing Systems
                   (NeurIPS 2023)",
  month         =  jun,
  year          =  2023,
  archivePrefix = "arXiv",
  primaryClass  = "q-bio.QM"
}

@article{hsu2022learning,
  title={Learning protein fitness models from evolutionary and assay-labeled data},
  author={Hsu, Chloe and Nisonoff, Hunter and Fannjiang, Clara and Listgarten, Jennifer},
  journal={Nature biotechnology},
  volume={40},
  number={7},
  pages={1114--1122},
  year={2022},
  publisher={Nature Publishing Group US New York}
}

@ARTICLE{Rapp2024-SAMPLE,
  title     = "Self-driving laboratories to autonomously navigate the protein
               fitness landscape",
  author    = "Rapp, Jacob T and Bremer, Bennett J and Romero, Philip A",
  journal   = "Nat Chem Eng",
  publisher = "Nature Publishing Group",
  volume    =  1,
  number    =  1,
  pages     = "97--107",
  month     =  jan,
  year      =  2024,
  language  = "en"
}

@ARTICLE{Yang2025-DE,
  title     = "Active learning-assisted directed evolution",
  author    = "Yang, Jason and Lal, Ravi G and Bowden, James C and Astudillo,
               Raul and Hameedi, Mikhail A and Kaur, Sukhvinder and Hill,
               Matthew and Yue, Yisong and Arnold, Frances H",
  journal   = "Nat. Commun.",
  publisher = "Springer Science and Business Media LLC",
  volume    =  16,
  number    =  1,
  pages     =  714,
  month     =  jan,
  year      =  2025,
  language  = "en"
}

@ARTICLE{Jiang2025-EVOLVEpro,
  title     = "Rapid in silico directed evolution by a protein language model
               with {EVOLVEpro}",
  author    = "Jiang, Kaiyi and Yan, Zhaoqing and Di Bernardo, Matteo and
               Sgrizzi, Samantha R and Villiger, Lukas and Kayabolen, Alisan and
               Kim, B J and Carscadden, Josephine K and Hiraizumi, Masahiro and
               Nishimasu, Hiroshi and Gootenberg, Jonathan S and Abudayyeh, Omar
               O",
  journal   = "Science",
  publisher = "American Association for the Advancement of Science",
  volume    =  387,
  number    =  6732,
  pages     = "eadr6006",
  month     =  jan,
  year      =  2025,
  language  = "en"
}

@article{biswas2021low,
  title={Low-N protein engineering with data-efficient deep learning},
  author={Biswas, Surojit and Khimulya, Grigory and Alley, Ethan C and Esvelt, Kevin M and Church, George M},
  journal={Nature methods},
  volume={18},
  number={4},
  pages={389--396},
  year={2021},
  publisher={Nature Publishing Group US New York}
}

@article{notin2023proteinnpt,
  title={Proteinnpt: Improving protein property prediction and design with non-parametric transformers},
  author={Notin, Pascal and Weitzman, Ruben and Marks, Debora and Gal, Yarin},
  journal={Advances in Neural Information Processing Systems},
  volume={36},
  pages={33529--33563},
  year={2023}
}

@article{notin2024proteingym,
  title={Proteingym: Large-scale benchmarks for protein fitness prediction and design},
  author={Notin, Pascal and Kollasch, Aaron and Ritter, Daniel and Van Niekerk, Lood and Paul, Steffanie and Spinner, Han and Rollins, Nathan and Shaw, Ada and Orenbuch, Rose and Weitzman, Ruben and others},
  journal={Advances in Neural Information Processing Systems},
  volume={36},
  year={2024}
}

@inproceedings{hawkins2024likelihood,
  title={Likelihood-based fine-tuning of protein language models for few-shot fitness prediction and design},
  author={Hawkins-Hooker, Alex and Kmec, Jakub and Bent, Oliver and Duckworth, Paul},
  booktitle={ICML 2024 Workshop on Efficient and Accessible Foundation Models for Biological Discovery},
  year={2024}
}

@article{beck2024metalic,
  title={Metalic: Meta-Learning In-Context with Protein Language Models},
  author={Beck, Jacob and Surana, Shikha and McAuliffe, Manus and Bent, Oliver and Barrett, Thomas D and Luis, Juan Jose Garau and Duckworth, Paul},
  journal={arXiv preprint arXiv:2410.08355},
  year={2024}
}

@article{nijkamp2023progen2,
  title={Progen2: exploring the boundaries of protein language models},
  author={Nijkamp, Erik and Ruffolo, Jeffrey A and Weinstein, Eli N and Naik, Nikhil and Madani, Ali},
  journal={Cell systems},
  volume={14},
  number={11},
  pages={968--978},
  year={2023},
  publisher={Elsevier}
}

@article{meier2021language,
  title={Language models enable zero-shot prediction of the effects of mutations on protein function},
  author={Meier, Joshua and Rao, Roshan and Verkuil, Robert and Liu, Jason and Sercu, Tom and Rives, Alex},
  journal={Advances in neural information processing systems},
  volume={34},
  pages={29287--29303},
  year={2021}
}

@ARTICLE{Notin2022-TranceptEVE,
  title    = "{TranceptEVE}: Combining family-specific and family-agnostic
              models of protein sequences for improved fitness prediction",
  author   = "Notin, Pascal and Van Niekerk, Lood and Kollasch, Aaron W and
              Ritter, Daniel and Gal, Yarin and Marks, Debora S",
  journal  = "36th Conference on Neural Information Processing Systems (NeurIPS 2022), LMRL workshop",
  pages    = "2022.12.07.519495",
  month    =  dec,
  year     =  2022,
  language = "en"
}

@inproceedings{notin2022tranception,
  title={Tranception: protein fitness prediction with autoregressive transformers and inference-time retrieval},
  author={Notin, Pascal and Dias, Mafalda and Frazer, Jonathan and Marchena-Hurtado, Javier and Gomez, Aidan N and Marks, Debora and Gal, Yarin},
  booktitle={International Conference on Machine Learning},
  pages={16990--17017},
  year={2022},
  organization={PMLR}
}

@InProceedings{pmlr-v119-sun20b,
  title = 	 {Test-Time Training with Self-Supervision for Generalization under Distribution Shifts},
  author =       {Sun, Yu and Wang, Xiaolong and Liu, Zhuang and Miller, John and Efros, Alexei and Hardt, Moritz},
  booktitle = 	 {Proceedings of the 37th International Conference on Machine Learning},
  pages = 	 {9229--9248},
  year = 	 {2020},
  editor = 	 {III, Hal Daumé and Singh, Aarti},
  volume = 	 {119},
  series = 	 {Proceedings of Machine Learning Research},
  month = 	 {13--18 Jul},
  publisher =    {PMLR},
  url = 	 {https://proceedings.mlr.press/v119/sun20b.html}
}

@inproceedings{
gandelsman2022testtime,
title={Test-Time Training with Masked Autoencoders},
author={Yossi Gandelsman and Yu Sun and Xinlei Chen and Alexei A Efros},
booktitle={Advances in Neural Information Processing Systems},
editor={Alice H. Oh and Alekh Agarwal and Danielle Belgrave and Kyunghyun Cho},
year={2022},
url={https://openreview.net/forum?id=SHMi1b7sjXk}
}

@misc{akyürek2024surprisingeffectivenesstesttimetraining,
      title={The Surprising Effectiveness of Test-Time Training for Abstract Reasoning}, 
      author={Ekin Akyürek and Mehul Damani and Linlu Qiu and Han Guo and Yoon Kim and Jacob Andreas},
      year={2024},
      eprint={2411.07279},
      archivePrefix={arXiv},
      primaryClass={cs.AI},
      url={https://arxiv.org/abs/2411.07279}, 
}

@inproceedings{
groth2024kermut,
title={Kermut: Composite kernel regression for protein variant effects},
author={Peter M{\o}rch Groth and Mads Herbert Kerrn and Lars Olsen and Jesper Salomon and Wouter Boomsma},
booktitle={The Thirty-eighth Annual Conference on Neural Information Processing Systems},
year={2024},
url={https://openreview.net/forum?id=jM9atrvUii}
}

@article{lin2023evolutionary,
  title={Evolutionary-scale prediction of atomic-level protein structure with a language model},
  author={Lin, Zeming and Akin, Halil and Rao, Roshan and Hie, Brian and Zhu, Zhongkai and Lu, Wenting and Smetanin, Nikita and Verkuil, Robert and Kabeli, Ori and Shmueli, Yaniv and others},
  journal={Science},
  volume={379},
  number={6637},
  pages={1123--1130},
  year={2023},
  publisher={American Association for the Advancement of Science}
}

@article{dai2020funnel,
  title={Funnel-transformer: Filtering out sequential redundancy for efficient language processing},
  author={Dai, Zihang and Lai, Guokun and Yang, Yiming and Le, Quoc},
  journal={Advances in neural information processing systems},
  volume={33},
  pages={4271--4282},
  year={2020}
}

@article{Notin2024MachineLF,
  title={Machine learning for functional protein design},
  author={Pascal Notin and Nathan J. Rollins and Yarin Gal and Chris Sander and Debora Marks},
  journal={Nature Biotechnology},
  year={2024},
  volume={42},
  pages={216-228},
  url={https://api.semanticscholar.org/CorpusID:267682799}
}

@article{Fowler2014DeepMS,
  title={Deep mutational scanning: a new style of protein science},
  author={Douglas M. Fowler and Stanley Fields},
  journal={Nature Methods},
  year={2014},
  volume={11},
  pages={801-807},
  url={https://api.semanticscholar.org/CorpusID:205422975}
}

@article{Tsuboyama2023MegascaleEA,
  title={Mega-scale experimental analysis of protein folding stability in biology and design},
  author={Kotaro Tsuboyama and Justas Dauparas and Jonathan Chen and {\'E}lodie Laine and Yasser Mohseni Behbahani and Jonathan J Weinstein and Niall M. Mangan and Sergey Ovchinnikov and Gabriel J. Rocklin},
  journal={Nature},
  year={2023},
  volume={620},
  pages={434 - 444},
  url={https://api.semanticscholar.org/CorpusID:259994520}
}

@article{Beltran2024SitesaturationMO,
  title={Site-saturation mutagenesis of 500 human protein domains},
  author={Antoni Beltran and Xianger Jiang and Yue Shen and Ben Lehner},
  journal={Nature},
  year={2024},
  volume={637},
  pages={885 - 894},
  url={https://api.semanticscholar.org/CorpusID:269502813}
}

@article{Somermeyer2021HeterogeneityOT,
  title={Heterogeneity of the GFP fitness landscape and data-driven protein design},
  author={Louisa Gonzalez Somermeyer and Aubin Fleiss and Alexander S. Mishin and Nina G. Bozhanova and Anna A. Igolkina and Jens Meiler and Maria-Elisenda Alaball Pujol and Ekaterina V. Putintseva and Karen S. Sarkisyan and Fyodor A. Kondrashov},
  journal={eLife},
  year={2021},
  volume={11},
  url={https://api.semanticscholar.org/CorpusID:245026524}
}

@article{Russ2020AnEM,
  title={An evolution-based model for designing chorismate mutase enzymes},
  author={William P. Russ and Matteo Figliuzzi and Christian Stocker and Pierre Barrat-Charlaix and Michael Socolich and Peter Kast and Donald Hilvert and R{\'e}mi Monasson and Simona Cocco and Martin Weigt and Rama Ranganathan},
  journal={Science},
  year={2020},
  volume={369},
  pages={440 - 445},
  url={https://api.semanticscholar.org/CorpusID:220714458}
}

@article{Suzek2014UniRefCA,
  title={UniRef clusters: a comprehensive and scalable alternative for improving sequence similarity searches},
  author={Baris E. Suzek and Yuqi Wang and Hongzhan Huang and Peter B. McGarvey and Cathy H. Wu},
  journal={Bioinformatics},
  year={2014},
  volume={31},
  pages={926 - 932},
  url={https://api.semanticscholar.org/CorpusID:12423917}
}

@article{Richardson2022MGnifyTM,
  title={MGnify: the microbiome sequence data analysis resource in 2023},
  author={Lorna J. Richardson and Ben Allen and Germana Baldi and Martin Beracochea and Maxwell L. Bileschi and Tony Burdett and Josephine Burgin and Juan Caballero and Guy Cochrane and Lucy J. Colwell and Tom Curtis and Alejandra Escobar-Zepeda and Tatiana A. Gurbich and Varsha Kale and Anton I. Korobeynikov and Shriya Raj and Alexander B. Rogers and Ekaterina A. Sakharova and Santiago Sanchez and Darren J. Wilkinson and Robert D. Finn},
  journal={Nucleic Acids Research},
  year={2022},
  volume={51},
  pages={D753 - D759},
  url={https://api.semanticscholar.org/CorpusID:254447132}
}

@article{ektefaie2024evaluating,
  title={Evaluating generalizability of artificial intelligence models for molecular datasets},
  author={Ektefaie, Yasha and Shen, Andrew and Bykova, Daria and Marin, Maximillian G and Zitnik, Marinka and Farhat, Maha},
  journal={Nature Machine Intelligence},
  volume={6},
  number={12},
  pages={1512--1524},
  year={2024},
  publisher={Nature Publishing Group}
}

@article{teufel2023graphpart,
  title={GraphPart: homology partitioning for biological sequence analysis},
  author={Teufel, Felix and G{\'\i}slason, Magn{\'u}s Halld{\'o}r and Almagro Armenteros, Jos{\'e} Juan and Johansen, Alexander Rosenberg and Winther, Ole and Nielsen, Henrik},
  journal={NAR genomics and bioinformatics},
  volume={5},
  number={4},
  pages={lqad088},
  year={2023},
  publisher={Oxford University Press}
}

@article{bernett2024guiding,
  title={Guiding questions to avoid data leakage in biological machine learning applications},
  author={Bernett, Judith and Blumenthal, David B and Grimm, Dominik G and Haselbeck, Florian and Joeres, Roman and Kalinina, Olga V and List, Markus},
  journal={Nature Methods},
  volume={21},
  number={8},
  pages={1444--1453},
  year={2024},
  publisher={Nature Publishing Group US New York}
}

@article{su2024roformer,
  title={Roformer: Enhanced transformer with rotary position embedding},
  author={Su, Jianlin and Ahmed, Murtadha and Lu, Yu and Pan, Shengfeng and Bo, Wen and Liu, Yunfeng},
  journal={Neurocomputing},
  volume={568},
  pages={127063},
  year={2024},
  publisher={Elsevier}
}

@article{hopf2017mutation,
   author = {Hopf, Thomas A and Ingraham, John B and Poelwijk, Frank J and Sch{\"a}rfe, Charlotta PI and Springer, Michael and Sander, Chris and Marks, Debora S},
   title = {Mutation effects predicted from sequence co-variation},
   journal = {Nat. Biotechnol.},
   volume = {35},
   number = {2},
   pages = {128},
   ISSN = {1546-1696},
   year = {2017},
   type = {Journal Article}
}

@article{Frazer2021DiseaseVP,
  title={Disease variant prediction with deep generative models of evolutionary data.},
  author={Jonathan Frazer and Pascal Notin and Mafalda Dias and Aidan Gomez and Joseph K Min and Kelly P. Brock and Yarin Gal and Debora S. Marks},
  journal={Nature},
  year={2021}
}

@misc{dong2024surveyincontextlearning,
      title={A Survey on In-context Learning}, 
      author={Qingxiu Dong and Lei Li and Damai Dai and Ce Zheng and Jingyuan Ma and Rui Li and Heming Xia and Jingjing Xu and Zhiyong Wu and Tianyu Liu and Baobao Chang and Xu Sun and Lei Li and Zhifang Sui},
      year={2024},
      eprint={2301.00234},
      archivePrefix={arXiv},
      primaryClass={cs.CL},
      url={https://arxiv.org/abs/2301.00234}, 
}

@article{hollmann2025accurate,
  title={Accurate predictions on small data with a tabular foundation model},
  author={Hollmann, Noah and M{\"u}ller, Samuel and Purucker, Lennart and Krishnakumar, Arjun and K{\"o}rfer, Max and Hoo, Shi Bin and Schirrmeister, Robin Tibor and Hutter, Frank},
  journal={Nature},
  volume={637},
  number={8045},
  pages={319--326},
  year={2025},
  publisher={Nature Publishing Group UK London}
}

@inproceedings{
gordon2025protein,
title={Protein Language Model Fitness is a Matter of Preference},
author={Cade W Gordon and Amy X. Lu and Pieter Abbeel},
booktitle={The Thirteenth International Conference on Learning Representations},
year={2025},
url={https://openreview.net/forum?id=UvPdpa4LuV}
}

@article{Plesa2018,
  author = {Plesa, Calin and Sidore, Angus M. and Lubock, Nathan B. and Zhang, Di and Kosuri, Sriram},
  title = {Multiplexed gene synthesis in emulsions for exploring protein functional landscapes},
  journal = {Science},
  volume = {359},
  number = {6373},
  pages = {343--347},
  year = {2018},
  month = jan,
  doi = {10.1126/science.aao5167},
  publisher = {American Association for the Advancement of Science},
  pmid = {29301959},
  pmc = {PMC6261299}
}

@article{Davidi2020HighlyAR,
  title={Highly active rubiscos discovered by systematic interrogation of natural sequence diversity},
  author={Dan Davidi and Melina Shamshoum and Zhijun Guo and Y. Bar-On and Noam Prywes and Aia Oz and Jagoda Jabłońska and Avi I. Flamholz and David G. Wernick and Niv Antonovsky and Benoit de Pins and Lior Shachar and Dina Hochhauser and Yoav Peleg and Shira Albeck and Itai Sharon and Oliver Mueller-Cajar and Ron Milo},
  journal={The EMBO Journal},
  year={2020},
  volume={39},
  url={https://api.semanticscholar.org/CorpusID:219327380}
}
\bibliographystyle{unsrt}

\newpage
\appendix

\label{sec:appendix}

 \setcounter{table}{0}
\renewcommand{\thetable}{A\arabic{table}}%
\setcounter{figure}{0}
\renewcommand{\thefigure}{A\arabic{figure}}%
\setcounter{equation}{0}
\renewcommand{\theequation}{A\arabic{equation}}%
\FloatBarrier
\section{Few-shot benchmark DMS assay selection}

We select a representative collection of assays from ProteinGym for benchmarking few-shot prediction performance. As opposed to previous selections, such as the one used by ProteinNPT and Metalic, we also ensure that the assays are reasonably independent with respect to the training data, so that no close homolog or identical protein was trained on.

We avoid benchmarking on assays that reported abundance, as we consider it an ill-defined target with less relevance for real-world protein optimization efforts.
Abundance, as measured by DMS, could be understood as a convolution of biological processes such as expression, stability and degradation. In few-shot protein engineering campaigns, these properties would typically be evaluated one-by-one in real units. However, we still consider abundance assays to be a useful data resource for pre-training on diverse experiments. 

We excluded both HIS7\_YEAST\_Pokusaeva\_2019 and CAPSD\_AAV2S\_Sinai\_2021 as their scale proved prohibitive for experimentation given our available GPU resources.

\begin{table}[!b]
\caption{The hold out set of ProteinGym DMS assays. Similarity to train is computed as pairwise Needleman-Wunsch global sequence identity of the wild type proteins.}
\label{tab:hold_out}
\vskip 0.15in
\begin{center}
\begin{small}
\begin{tabular}{lcccr}
\toprule
Assay & Type & Similarity to train & Closest protein\\
\midrule
AMFR\_HUMAN\_Tsuboyama\_2023\_4G3O & Stability & 23\% & CUE1\_YEAST\\
RCD1\_ARATH\_Tsuboyama\_2023\_5OAO & Stability & 20\% & NUSG\_MYCTU\\
SR43C\_ARATH\_Tsuboyama\_2023\_2N88 & Stability & 39\% & CBX4\_HUMAN\\
FECA\_ECOLI\_Tsuboyama\_2023\_2D1U & Stability & 20\% & RPC1\_BP434\\
PKN1\_HUMAN\_Tsuboyama\_2023\_1URF & Stability & 21\% & DN7A\_SACS2\\
CSN4\_MOUSE\_Tsuboyama\_2023\_1UFM & Stability & 20\% & UBE4B\_HUMAN\\
SPA\_STAAU\_Tsuboyama\_2023\_1LP1 & Stability & 22\%& HECD1\_HUMAN\\
NKX31\_HUMAN\_Tsuboyama\_2023\_2L9R & Stability & 32\% & PITX2\_HUMAN\\
EPHB2\_HUMAN\_Tsuboyama\_2023\_1F0M & Stability & 25\%& PR40A\_HUMAN\\
SQSTM\_MOUSE\_Tsuboyama\_2023\_2RRU & Stability & 29\% &OTU7A\_HUMAN\\
MAFG\_MOUSE\_Tsuboyama\_2023\_1K1V & Stability & 24\% & RPB1\_HUMAN\\
SCIN\_STAAR\_Tsuboyama\_2023\_2QFF & Stability &23\%& HVP\_LAMBD\\
DNJA1\_HUMAN\_Tsuboyama\_2023\_2LO1 & Stability&23\%&HECD1\_HUMAN\\
VRPI\_BPT7\_Tsuboyama\_2023\_2WNM & Stability&19\%&MYO3\_YEAST\\
ESTA\_BACSU\_Nutschel\_2020 & Stability&15\%&CALM1\_HUMAN\\
CASP3\_HUMAN\_Roychowdhury\_2020 & Enz. Activity&48\%&CASP7\_HUMAN\\
BLAT\_ECOLX\_Deng\_2012 & Enz. Activity&18\%&CD19\_HUMAN\\
BLAT\_ECOLX\_Jacquier\_2013 & Enz. Activity&18\%&CD19\_HUMAN\\
BLAT\_ECOLX\_Stiffler\_2015 & Enz. Activity&18\%&CD19\_HUMAN\\
BLAT\_ECOLX\_Firnberg\_2014 & Enz. Activity&18\%&CD19\_HUMAN\\
VKOR1\_HUMAN\_Chiasson\_2020\_activity & Enz. Activity&13\%&RPC1\_BP434\\
VKOR1\_HUMAN\_Chiasson\_2020\_abundance & Abundance&13\%&RPC1\_BP434\\
Q8WTC7\_9CNID\_Somermeyer\_2022 & Fluoresence &18\%&Q6WV13\_9MAXI\\
D7PM05\_CLYGR\_Somermeyer\_2022 & Fluoresence&19\%&MTH3\_HAEAE\\
GFP\_AEQVI\_Sarkisyan\_2016 & Fluoresence&18\%&Q6WV13\_9MAXI\\
DLG4\_RAT\_McLaughlin\_2012 & Binding&19\%& PSAE\_SYNP2\\
RL40A\_YEAST\_Roscoe\_2013 & Binding&20\%&SPG2\_STRSG\\
GRB2\_HUMAN\_Faure\_2021 & Binding&27\%& SRBS1\_HUMAN\\
DYR\_ECOLI\_Thompson\_2019& Enz. Activity&15\%&NUD15\_HUMAN\\
DLG4\_HUMAN\_Faure\_2021 & Binding &25\%& EPHB2\_HUMAN\\
RL40A\_YEAST\_Mavor\_2016 & Binding&20\%&SPG2\_STRSG\\
DYR\_ECOLI\_Nguyen\_2023& Enz. Activity&15\%&NUD15\_HUMAN \\
RL40A\_YEAST\_Roscoe\_2014& Binding&20\%&SPG2\_STRSG\\

\bottomrule
\end{tabular}
\end{small}
\end{center}
\vskip -0.1in
\end{table}

\FloatBarrier

\begin{longtable}[!b]{lc}
\label{tab:training_data} \\
\caption{The training set of ProteinGym DMS assays.}\\
\toprule
Assay & Type\\
\midrule
\endfirsthead
Assay & Type\\
\midrule
\endhead
A4GRB6\_PSEAI\_Chen\_2020 & Enz. Activity \\
AACC1\_PSEAI\_Dandage\_2018 & Enz. Activity \\
ACE2\_HUMAN\_Chan\_2020 & Binding \\
AICDA\_HUMAN\_Gajula\_2014\_3cycles & Enz. Activity \\
AMIE\_PSEAE\_Wrenbeck\_2017 & Enz. Activity \\
ANCSZ\_Hobbs\_2022 & Enz. Activity \\
ARGR\_ECOLI\_Tsuboyama\_2023\_1AOY & Stability \\
B2L11\_HUMAN\_Dutta\_2010\_binding-Mcl-1 & Binding \\
BBC1\_YEAST\_Tsuboyama\_2023\_1TG0 & Stability \\
BCHB\_CHLTE\_Tsuboyama\_2023\_2KRU & Stability \\
CALM1\_HUMAN\_Weile\_2017 & Binding \\
CAS9\_STRP1\_Spencer\_2017\_positive & Enz. Activity \\
CASP7\_HUMAN\_Roychowdhury\_2020 & Enz. Activity \\
CATR\_CHLRE\_Tsuboyama\_2023\_2AMI & Stability \\
CBPA2\_HUMAN\_Tsuboyama\_2023\_1O6X & Stability \\
CBS\_HUMAN\_Sun\_2020 & Enz. Activity \\
CBX4\_HUMAN\_Tsuboyama\_2023\_2K28 & Stability \\
CD19\_HUMAN\_Klesmith\_2019\_FMC\_singles & Binding \\
CP2C9\_HUMAN\_Amorosi\_2021\_abundance & Abundance \\
CP2C9\_HUMAN\_Amorosi\_2021\_activity & Binding \\
CUE1\_YEAST\_Tsuboyama\_2023\_2MYX & Stability \\
DN7A\_SACS2\_Tsuboyama\_2023\_1JIC & Stability \\
DOCK1\_MOUSE\_Tsuboyama\_2023\_2M0Y & Stability \\
ENVZ\_ECOLI\_Ghose\_2023 & Enz. Activity \\
ERBB2\_HUMAN\_Elazar\_2016 & Abundance \\
F7YBW8\_MESOW\_Aakre\_2015 & Binding \\
F7YBW8\_MESOW\_Ding\_2023 & Binding \\ 
FKBP3\_HUMAN\_Tsuboyama\_2023\_2KFV & Stability \\
GDIA\_HUMAN\_Silverstein\_2021 & Binding \\
GLPA\_HUMAN\_Elazar\_2016 & Abundance \\
HCP\_LAMBD\_Tsuboyama\_2023\_2L6Q & Stability \\
HECD1\_HUMAN\_Tsuboyama\_2023\_3DKM & Stability \\
HMDH\_HUMAN\_Jiang\_2019 & Enz. Activity \\
HXK4\_HUMAN\_Gersing\_2022\_activity & Enz. Activity \\
HXK4\_HUMAN\_Gersing\_2023\_abundance & Abundance \\
ILF3\_HUMAN\_Tsuboyama\_2023\_2L33 & Stability \\
ISDH\_STAAW\_Tsuboyama\_2023\_2LHR & Stability \\
KKA2\_KLEPN\_Melnikov\_2014 & Enz. Activity \\
LGK\_LIPST\_Klesmith\_2015 & Enz. Activity \\
LYAM1\_HUMAN\_Elazar\_2016 & Abundance \\
MBD11\_ARATH\_Tsuboyama\_2023\_6ACV & Stability \\
MET\_HUMAN\_Estevam\_2023 & Enz. Activity \\
MK01\_HUMAN\_Brenan\_2016 & Enz. Activity \\
MSH2\_HUMAN\_Jia\_2020 & Enz. Activity \\
MTH3\_HAEAE\_RockahShmuel\_2015 & Enz. Activity \\
MTHR\_HUMAN\_Weile\_2021 & Enz. Activity \\
MYO3\_YEAST\_Tsuboyama\_2023\_2BTT & Stability \\
NUD15\_HUMAN\_Suiter\_2020 & Enz. Activity \\
NUSA\_ECOLI\_Tsuboyama\_2023\_1WCL & Stability \\
NUSG\_MYCTU\_Tsuboyama\_2023\_2MI6 & Stability \\
OBSCN\_HUMAN\_Tsuboyama\_2023\_1V1C & Stability \\
ODP2\_GEOSE\_Tsuboyama\_2023\_1W4G & Stability \\
OPSD\_HUMAN\_Wan\_2019 & Abundance \\
OTC\_HUMAN\_Lo\_2023 & Enz. Activity \\
OTU7A\_HUMAN\_Tsuboyama\_2023\_2L2D & Stability \\
OXDA\_RHOTO\_Vanella\_2023\_activity & Enz. Activity \\
P84126\_THETH\_Chan\_2017 & Enz. Activity \\
PAI1\_HUMAN\_Huttinger\_2021 & Binding \\
PIN1\_HUMAN\_Tsuboyama\_2023\_1I6C & Stability \\
PITX2\_HUMAN\_Tsuboyama\_2023\_2L7M & Stability \\
POLG\_PESV\_Tsuboyama\_2023\_2MXD & Stability \\
PPARG\_HUMAN\_Majithia\_2016 & Binding \\
PR40A\_HUMAN\_Tsuboyama\_2023\_1UZC & Stability \\
PRKN\_HUMAN\_Clausen\_2023 & Abundance \\
PSAE\_PICP2\_Tsuboyama\_2023\_1PSE & Abundance \\
PTEN\_HUMAN\_Mighell\_2018 & Enz. Activity \\
Q53Z42\_HUMAN\_McShan\_2019\_binding-TAPBPR & Binding \\
Q59976\_STRSQ\_Romero\_2015 & Enz. Activity \\
Q6WV12\_9MAXI\_Somermeyer\_2022 & Fluorescence \\
RAD\_ANTMA\_Tsuboyama\_2023\_2CJJ & Stability \\
RAF1\_HUMAN\_Zinkus-Boltz\_2019 & Binding \\
RASH\_HUMAN\_Bandaru\_2017 & Enz. Activity \\
RASK\_HUMAN\_Weng\_2022\_abundance & Abundance \\
RASK\_HUMAN\_Weng\_2022\_binding-DARPin\_K55 & Binding \\
RBP1\_HUMAN\_Tsuboyama\_2023\_2KWH & Stability \\
RCRO\_LAMBD\_Tsuboyama\_2023\_1ORC & Stability \\
RD23A\_HUMAN\_Tsuboyama\_2023\_1IFY & Stability \\
RFAH\_ECOLI\_Tsuboyama\_2023\_2LCL & Stability \\
RL20\_AQUAE\_Tsuboyama\_2023\_1GYZ & Stability \\
RNC\_ECOLI\_Weeks\_2023 & Enz. Activity \\
RPC1\_BP434\_Tsuboyama\_2023\_1R69 & Stability \\
RS15\_GEOSE\_Tsuboyama\_2023\_1A32 & Stability \\
SAV1\_MOUSE\_Tsuboyama\_2023\_2YSB & Stability \\
SBI\_STAAM\_Tsuboyama\_2023\_2JVG & Stability \\
SDA\_BACSU\_Tsuboyama\_2023\_1PV0 & Stability \\
SERC\_HUMAN\_Xie\_2023 & Enz. Activity \\
SHOC2\_HUMAN\_Kwon\_2022 & Binding \\
SOX30\_HUMAN\_Tsuboyama\_2023\_7JJK & Stability \\
SPG1\_STRSG\_Olson\_2014 & Binding \\
SPG1\_STRSG\_Wu\_2016 & Binding \\
SPG2\_STRSG\_Tsuboyama\_2023\_5UBS & Stability \\
SPIKE\_SARS2\_Starr\_2020\_binding & Binding \\
SPIKE\_SARS2\_Starr\_2020\_expression & Abundance \\
SPTN1\_CHICK\_Tsuboyama\_2023\_1TUD & Stability \\
SRBS1\_HUMAN\_Tsuboyama\_2023\_2O2W & Stability \\
SRC\_HUMAN\_Ahler\_2019 & Enz. Activity \\
SRC\_HUMAN\_Chakraborty\_2023\_binding-DAS\_25uM & Enz. Activity \\
TCRG1\_MOUSE\_Tsuboyama\_2023\_1E0L & Stability \\
THO1\_YEAST\_Tsuboyama\_2023\_2WQG & Stability \\
TNKS2\_HUMAN\_Tsuboyama\_2023\_5JRT & Stability \\
TPK1\_HUMAN\_Weile\_2017 & Enz. Activity \\
TPMT\_HUMAN\_Matreyek\_2018 & Abundance \\
UBC9\_HUMAN\_Weile\_2017 & Enz. Activity \\
UBE4B\_HUMAN\_Tsuboyama\_2023\_3L1X & Stability \\
UBE4B\_MOUSE\_Starita\_2013 & Enz. Activity \\
UBR5\_HUMAN\_Tsuboyama\_2023\_1I2T & Stability \\
VG08\_BPP22\_Tsuboyama\_2023\_2GP8 & Stability \\
VILI\_CHICK\_Tsuboyama\_2023\_1YU5 & Stability \\
YAIA\_ECOLI\_Tsuboyama\_2023\_2KVT & Stability \\
YAP1\_HUMAN\_Araya\_2012 & Binding \\
YNZC\_BACSU\_Tsuboyama\_2023\_2JVD & Stability \\

\bottomrule
\end{longtable}

\FloatBarrier

\section{Experimental details}


We follow ProteinNPT \cite{notin2023proteinnpt} for the Ridge regression baseline. Specifically, we process the mean-pooled ESM embedding using a linear layer, and the zero-shot score from ProGen using another linear layer without bias which is initialized with weight 1.0. We apply an L2 penalty of $5 \times 10^{-3}$ for the embedding linear layer and $1 \times 10^{-8}$ for the zero-shot linear layer. The models are trained for 1500 steps at a learning rate of 0.01. The Gaussian process (GP) model was adapted from Kermut \citep{groth2024kermut}. We reuse model components and the training loop from the official Kermut code release, omitting the structure kernel from the GP's composite kernel. GP models are trained for 150 steps at a learning rate of $3 \times 10^{-4}$.
The random forest (RF) baseline including all its hyperparameters was taken from the official EvolvePro codebase \citep{Jiang2025-EVOLVEpro}.


\begin{table}[H]
    \centering
    \caption{Hyperparameters of PRIMO. Unless stated otherwise, the architectural configuration of the transformer is based on the ESM-2 attention block, with the PRIMO layer adaptations as discussed in the main text. For the TTT phase, we only list parameters that differ from the pre-training setup. PRIMO was trained on a single RTX 6000, using gradient accumulation to enable the specified batch size.}
    \vskip 0.15in
    \begin{tabular}{lc}
    \toprule
    Parameter & Value \\
    \midrule
    Layers     &  6\\
    Hidden size     &  400\\
    Attention heads     & 8 \\
    Feedforward factor & 4 \\
    Pooling vectors per sequence & 3\\
    AA Embedder & ESM-2 650M \cite{lin2023evolutionary} \\
    Zero-shot predictor & ProGen-2 medium \cite{nijkamp2023progen2} \\
    \midrule
    Dropout & 0.1 \\
    Weight decay & 0.01 \\
    Gradient norm clipping & 1.0 \\
    Learning rate & 0.0001\\
    Learning rate schedule & Triangular \\
    Warmup steps & 1000 \\
    Total sets & 150,000 \\
    Set size & 32 \\
    Batch size & 12 \\
    AA sequence length & 512 \\
    Start MLM loss factor & 0.5 \\
    End MLM loss factor & 0.05 \\
    MLM loss factor schedule & Cosine \\
    \midrule
    Label masking probability & 0.33 \\
    Span masking probability\tablefootnote{The span masking probability is applied after the label masking probability, so that the effective probability becomes $(1-0.33)*0.2$.} & 0.2 \\
    Minimum span fraction & 0.05 \\
    Maximum span fraction & 0.15 \\
    Span length distribution & Uniform \\
    Mask all variant positions\tablefootnote{This means that the sampled span length will be split over $k$ mutation positions as needed.} & True \\
    \midrule 
    TTT learning rate & 0.0001 \\
    TTT learning rate schedule & Flat \\
    TTT training set size & min($N$, 32) \\
    TTT sequence length & $L$ \\
    TTT total steps & 25 \\
    TTT/ICL inference set size & $ N+1$ \\
         \bottomrule
    \end{tabular}
    \label{tab:my_label}
\end{table}

\FloatBarrier

\section{Additional results}
\FloatBarrier

\subsection{Ablations}

We ablate the following components of PRIMO:
\begin{itemize}
    \item Conditioning approach for zero-shot prediction (\autoref{tab:reference_point})
    \item Loss function (\autoref{tab:ablation_loss})
    \item Auxiluary inputs (Property type and pLM zero-shot score) (\autoref{tab:input_ablation})
    \item TTT protocol (\autoref{tab:ttt_steps})
    \item Attention mechanism (\autoref{tab:ablation_attention})
\end{itemize}

For efficiency, ablation experiments are performed at a constant set size of $N=16$ (with the exception of the zero-shot ablation). Overall, the ablations demonstrate that the modeling choices of PRIMO are sensible, and all individually contribute to improved performance.

\begin{table}[!h]
    \centering
    \caption{Effect of providing an arbitrary reference datapoint with fitness set to 0.5 (midpoint of the min-max scale) for calibration when performing 0-shot prediction with PRIMO. When no context sequence is provided, the prediction performance of the logits returned by PRIMO collapses.}
    \label{tab:reference_point}
    \vskip 0.15in
\begin{tabular}{lr}
    \toprule
        Approach & Average Spearman \\
    \midrule
        Unlabeled sequence only &  - 0.257\\
        Reference sequence with arbitrary fitness & 0.522 \\
    \bottomrule
    \end{tabular}
\end{table}

\begin{table}[!h]
    \centering
    \caption{Performance of PRIMO (TTT) when training using different loss functions at $N$=16. }
    \vskip 0.15in
    \label{tab:ablation_loss}
    \begin{tabular}{lc}
\toprule
Loss &	Average Spearman \\
\midrule
Preference &	0.538 \\
MSE	& 0.528 \\
\bottomrule
    \end{tabular}
\end{table}

\begin{table}[!h]
    \centering
    \caption{Performance of PRIMO (TTT) when ablating additional inputs at $N$=16. }
    \label{tab:input_ablation}
    \vskip 0.15in
    \label{ablation_loss}
    \begin{tabular}{lc}
\toprule
Inputs &	Average Spearman \\
\midrule
All &	0.538 \\
w/o Property ID	& 0.532 \\
w/o zero-shot score	& 0.529 \\
\bottomrule
    \end{tabular}
\end{table}

\begin{table}[!h]
    \centering
    \caption{Performance of PRIMO (TTT) with different numbers of TTT steps at $N$=16.}
    \label{tab:ttt_steps}
    \vskip 0.15in
    \begin{tabular}{lc}
\toprule
Steps	&Avg. Spearman\\
\midrule
0 (ICL)	& 0.513\\
10 &	0.526\\
25&	0.539\\
50	&0.526\\
75	&0.518\\
100	&0.515\\
\bottomrule
    \end{tabular}
\end{table}

\begin{table}[]
    \centering
    \caption{Performance of PRIMO models using different attention mechanisms at $N$=16. \textit{Tiered} refers to PoET's tiered attention,  \textit{Pooled} is the attention mechanism used by the final PRIMO model described in the main text. The tiered attention model was trained at a pre-training set size of 16 due to increased computational complexity. Results were obtained on a preliminary data split before training the final PRIMO model.}
    \vskip 0.15in
    \label{tab:ablation_attention}
    \begin{tabular}{lcc}

\toprule
 &	Tiered & Pooled \\
\midrule
ICL &	0.45 &	0.48 \\
TTT	&0.43	&0.47 \\
\bottomrule
    \end{tabular}
\end{table}

\FloatBarrier
\subsection{Assay level performance}

\begin{landscape}

\begin{table}[]
    \centering
    \caption{Per-assay results on the hold out set (1/2).}
    \label{tab:results_1}
    \vskip 0.15in
\begin{adjustbox}{max width=1.3\textwidth}

\begin{tabular}{lllllllllllllllllll}
 &  & \rotatebox{270}{AMFR\_HUMAN\_Tsuboyama\_2023\_4G3O} & \rotatebox{270}{BLAT\_ECOLX\_Deng\_2012} & \rotatebox{270}{BLAT\_ECOLX\_Firnberg\_2014} & 
 \rotatebox{270}{BLAT\_ECOLX\_Gonzalez\_2019} & \rotatebox{270}{BLAT\_ECOLX\_Jacquier\_2013} & \rotatebox{270}{BLAT\_ECOLX\_Stiffler\_2015} & \rotatebox{270}{CASP3\_HUMAN\_Roychowdhury\_2020} & \rotatebox{270}{CSN4\_MOUSE\_Tsuboyama\_2023\_1UFM} & \rotatebox{270}{D7PM05\_CLYGR\_Somermeyer\_2022} & \rotatebox{270}{DLG4\_HUMAN\_Faure\_2021} & \rotatebox{270}{DLG4\_RAT\_McLaughlin\_2012} & \rotatebox{270}{DNJA1\_HUMAN\_Tsuboyama\_2023\_2LO1} & \rotatebox{270}{DYR\_ECOLI\_Nguyen\_2023} & \rotatebox{270}{DYR\_ECOLI\_Thompson\_2019} & \rotatebox{270}{EPHB2\_HUMAN\_Tsuboyama\_2023\_1F0M} & \rotatebox{270}{ESTA\_BACSU\_Nutschel\_2020} & \rotatebox{270}{FECA\_ECOLI\_Tsuboyama\_2023\_2D1U} \\
 Shots & Method  &  &  &  &  &  &  &  &  &  &  &  &  &  &  &  &  &  \\
\midrule
\multirow[t]{2}{*}{0} & ProGen & 0.01 & 0.43 & 0.64 & 0.62 & 0.64 & 0.19 & 0.51 & 0.57 & 0.13 & 0.61 & 0.41 & 0.75 & 0.44 & 0.39 & 0.65 & 0.26 & 0.50\\
 & PRIMO (ICL) & 0.73 ± 0.03 & 0.44 ± 0.02 & 0.54 ± 0.17 & 0.56 ± 0.03 & 0.48 ± 0.25 & 0.67 ± 0.01 & 0.43 ± 0.21 & 0.66 ± 0.01 & 0.06 ± 0.0 & 0.58 ± 0.0 & 0.4 ± 0.0 & 0.84 ± 0.01 & 0.45 ± 0.06 & 0.41 ± 0.07 & 0.87 ± 0.01 & 0.24 ± 0.01 & 0.66 ± 0.01 \\
 & PRIMO (TTT) & 0.71 ± 0.02 & 0.42 ± 0.01 & 0.66 ± 0.01 & 0.57 ± 0.02 & 0.6 ± 0.01 & 0.65 ± 0.02 & 0.47 ± 0.14 & 0.59 ± 0.06 & 0.18 ± 0.01 & 0.53 ± 0.01 & 0.38 ± 0.0 & 0.8 ± 0.02 & 0.44 ± 0.06 & 0.44 ± 0.01 & 0.85 ± 0.02 & 0.27 ± 0.01 & 0.62 ± 0.03 \\
\midrule
\multirow[t]{5}{*}{4} & GP & 0.36 ± 0.24 & 0.12 ± 0.17 & 0.21 ± 0.13 & 0.13 ± 0.51 & 0.16 ± 0.11 & 0.17 ± 0.15 & 0.09 ± 0.18 & 0.21 ± 0.28 & -0.1 ± 0.05 & 0.27 ± 0.07 & 0.14 ± 0.25 & 0.61 ± 0.24 & -0.01 ± 0.13 & 0.04 ± 0.09 & 0.7 ± 0.04 & 0.02 ± 0.06 & 0.38 ± 0.07 \\
\bottomrule
\end{tabular}

\end{adjustbox}

\end{table}
\end{landscape}

\begin{landscape}

\begin{table}[]
    \centering
    \caption{Per-assay results on the hold out set (2/2).}
    \label{tab:results_2}
    \vskip 0.15in
\begin{adjustbox}{max width=1.3\textwidth}

\begin{tabular}{lllllllllllllllllll}

 &  & \rotatebox{270}{GFP\_AEQVI\_Sarkisyan\_2016} & \rotatebox{270}{GRB2\_HUMAN\_Faure\_2021} & \rotatebox{270}{MAFG\_MOUSE\_Tsuboyama\_2023\_1K1V} & \rotatebox{270}{NKX31\_HUMAN\_Tsuboyama\_2023\_2L9R} & \rotatebox{270}{PKN1\_HUMAN\_Tsuboyama\_2023\_1URF} & \rotatebox{270}{Q8WTC7\_9CNID\_Somermeyer\_2022} & \rotatebox{270}{RCD1\_ARATH\_Tsuboyama\_2023\_5OAO} & \rotatebox{270}{RL40A\_YEAST\_Mavor\_2016} & \rotatebox{270}{RL40A\_YEAST\_Roscoe\_2013} & \rotatebox{270}{RL40A\_YEAST\_Roscoe\_2014} & \rotatebox{270}{SCIN\_STAAR\_Tsuboyama\_2023\_2QFF} & \rotatebox{270}{SPA\_STAAU\_Tsuboyama\_2023\_1LP1} & \rotatebox{270}{SQSTM\_MOUSE\_Tsuboyama\_2023\_2RRU} & \rotatebox{270}{SR43C\_ARATH\_Tsuboyama\_2023\_2N88} & \rotatebox{270}{VKOR1\_HUMAN\_Chiasson\_2020\_abundance} & \rotatebox{270}{VKOR1\_HUMAN\_Chiasson\_2020\_activity} & \rotatebox{270}{VRPI\_BPT7\_Tsuboyama\_2023\_2WNM} \\
 & Shots &  &  &  &  &  &  &  &  &  &  &  &  &  &  &  &  &  \\
\midrule
\multirow[t]{2}{*}{0} & ProGen & 0.15 & 0.53 & 0.51 & 0.58 & 0.43 & 0.01 & 0.38 & 0.42 & 0.48 & .38 & 0.13 & -0.09 & 0.57 & 0.56 & 0.50 & 0.41 & 0.11\\
 & PRIMO (ICL) & 0.08 ± 0.0 & 0.6 ± 0.0 & 0.2 ± 0.04 & 0.75 ± 0.0 & 0.71 ± 0.0 & 0.05 ± 0.0 & 0.72 ± 0.0 & 0.4 ± 0.0 & 0.45 ± 0.0 & 0.54 ± 0.0 & 0.51 ± 0.01 & 0.2 ± 0.0 & 0.76 ± 0.01 & 0.78 ± 0.04 & 0.45 ± 0.0 & 0.35 ± 0.01 & 0.78 ± 0.07 \\
 & PRIMO (TTT) & 0.08 ± 0.05 & 0.59 ± 0.01 & 0.34 ± 0.07 & 0.65 ± 0.11 & 0.55 ± 0.02 & 0.08 ± 0.01 & 0.66 ± 0.04 & 0.38 ± 0.01 & 0.45 ± 0.02 & 0.47 ± 0.03 & 0.57 ± 0.02 & 0.26 ± 0.02 & 0.73 ± 0.02 & 0.81 ± 0.02 & 0.42 ± 0.01 & 0.34 ± 0.01 & 0.76 ± 0.01 \\
\bottomrule
\end{tabular}
\end{adjustbox}

\end{table}
\end{landscape}

\newpage

\subsection{PRIMO vs Metalic}
                        
\begin{table*}[h]
\caption{Performance of a PRIMO and Metalic model on substitution few-shot prediction of held-out DMS assays, using the PRIMO split.}
\label{tab:metalic_vs_primo}
\vskip 0.15in
\begin{center}
\begin{small}
\begin{tabular}{llllllll}
\toprule
Method                & \multicolumn{7}{l}{Shots}            \\
                      & 0  & 4  & 8 & 16   & 32   & 64   & 128  \\
\midrule
Overall\\
\midrule
PRIMO (ICL) & 0.43 ± 0.0 & 0.44 ± 0.0 & 0.44 ± 0.0 & 0.44 ± 0.0 & 0.44 ± 0.0 & 0.44 ± 0.0 & 0.45 ± 0.01 \\
PRIMO (TTT) & 0.46 ± 0.0 & 0.46 ± 0.01 & 0.47 ± 0.02 & 0.52 ± 0.02 & 0.56 ± 0.01 & 0.61 ± 0.01 & 0.66 ± 0.01 \\
Metalic & 0.34 ± 0.02 & 0.34 ± 0.01 & 0.38 ± 0.01 & 0.39 ± 0.01 & 0.43 ± 0.00 & 0.47 ± 0.00 & 0.54 ± 0.00 \\
\midrule
Stability\\
\midrule
PRIMO (ICL) & 0.58 ± 0.0 & 0.58 ± 0.0 & 0.59 ± 0.0 & 0.59 ± 0.0 & 0.59 ± 0.0 & 0.59 ± 0.0 & 0.59 ± 0.0 \\
\bottomrule
\end{tabular}
\end{small}
\end{center}
\end{table*}

\newpage

\subsection{Performance on Substitution and Indel}

\begin{table*}[h]
\caption{Substitution few-shot prediction of held-out DMS assays.}
\label{tab:substitution_averages}
\vskip 0.15in
\begin{center}
\begin{small}
\begin{tabular}{llllllll}
\toprule
Method                & \multicolumn{7}{l}{Shots}            \\
                      & 0  & 4  & 8 & 16   & 32   & 64   & 128  \\
\midrule
Overall\\
\midrule
GP & - & 0.25 ± 0.01 & 0.32 ± 0.02 & 0.38 ± 0.02 & 0.44 ± 0.01 & 0.5 ± 0.01 & 0.55 ± 0.0 \\
Ridge & - & 0.25 ± 0.02 & 0.32 ± 0.02 & 0.4 ± 0.01 & 0.48 ± 0.01 & 0.56 ± 0.01 & 0.63 ± 0.0 \\
RF & - & 0.22 ± 0.01 & 0.31 ± 0.01 & 0.38 ± 0.01 & 0.45 ± 0.01 & 0.52 ± 0.01 & 0.59 ± 0.0 \\
PRIMO (ICL) & 0.5 ± 0.01 & 0.53 ± 0.0 & 0.53 ± 0.0 & 0.53 ± 0.0 & 0.53 ± 0.0 & 0.53 ± 0.0 & 0.53 ± 0.0 \\
PRIMO (TTT) & 0.5 ± 0.01 & 0.48 ± 0.01 & 0.5 ± 0.01 & 0.53 ± 0.01 & 0.58 ± 0.02 & 0.62 ± 0.0 & 0.67 ± 0.01 \\
\bottomrule
\end{tabular}
\end{small}
\end{center}
\end{table*}

\begin{table*}[]
\caption{Indel few-shot prediction of held-out DMS assays.}
\label{tab:indel_averages}
\vskip 0.15in
\begin{center}
\begin{small}
\begin{tabular}{llllllll}
\toprule
Method                & \multicolumn{7}{l}{Shots}            \\
                      & 0  & 4  & 8 & 16   & 32   & 64   & 128  \\
\midrule
Overall/Stability\\
\midrule
GP  & - & 0.29 ± 0.09 & 0.43 ± 0.09 & 0.5 ± 0.02 & 0.54 ± 0.03 & 0.59 ± 0.02 & 0.64 ± 0.02 \\
Ridge & - & 0.42 ± 0.06 & 0.49 ± 0.07 & 0.56 ± 0.04 & 0.59 ± 0.02 & 0.63 ± 0.02 & 0.71 ± 0.03 \\
RF & - & 0.39 ± 0.05 & 0.48 ± 0.06 & 0.55 ± 0.05 & 0.56 ± 0.03 & 0.64 ± 0.02 & 0.69 ± 0.02 \\
PRIMO (ICL) & 0.64 ± 0.0 & 0.65 ± 0.01 & 0.65 ± 0.01 & 0.64 ± 0.0 & 0.64 ± 0.0 & 0.64 ± 0.0 & 0.64 ± 0.0 \\
PRIMO (TTT) & 0.65 ± 0.01 & 0.65 ± 0.02 & 0.66 ± 0.03 & 0.67 ± 0.01 & 0.69 ± 0.01 & 0.71 ± 0.01 & 0.75 ± 0.02 \\
\bottomrule
\end{tabular}
\end{small}
\end{center}
\end{table*}

\begin{landscape}
\begin{table}[]
    \centering
    \caption{Substititution per-assay results on the hold out set (1/2).}
    \label{tab:sub_results_1}
    \vskip 0.15in
\begin{adjustbox}{max width=1.6\textwidth}

\begin{tabular}{lllllllllllllllllll}
 &  & \rotatebox{270}{AMFR\_HUMAN\_Tsuboyama\_2023\_4G3O} & \rotatebox{270}{BLAT\_ECOLX\_Deng\_2012} & \rotatebox{270}{BLAT\_ECOLX\_Firnberg\_2014} & \rotatebox{270}{BLAT\_ECOLX\_Gonzalez\_2019} & \rotatebox{270}{BLAT\_ECOLX\_Jacquier\_2013} & \rotatebox{270}{BLAT\_ECOLX\_Stiffler\_2015} & \rotatebox{270}{CASP3\_HUMAN\_Roychowdhury\_2020} & \rotatebox{270}{CSN4\_MOUSE\_Tsuboyama\_2023\_1UFM} & \rotatebox{270}{D7PM05\_CLYGR\_Somermeyer\_2022} & \rotatebox{270}{DLG4\_HUMAN\_Faure\_2021} & \rotatebox{270}{DLG4\_RAT\_McLaughlin\_2012} & \rotatebox{270}{DNJA1\_HUMAN\_Tsuboyama\_2023\_2LO1} & \rotatebox{270}{DYR\_ECOLI\_Nguyen\_2023} & \rotatebox{270}{DYR\_ECOLI\_Thompson\_2019} & \rotatebox{270}{EPHB2\_HUMAN\_Tsuboyama\_2023\_1F0M} & \rotatebox{270}{ESTA\_BACSU\_Nutschel\_2020} & \rotatebox{270}{FECA\_ECOLI\_Tsuboyama\_2023\_2D1U} \\
 Shots & Method  &  &  &  &  &  &  &  &  &  &  &  &  &  &  &  &  &  \\
\midrule
\midrule
\multirow[t]{2}{*}{1} & PRIMO (ICL) & 0.73 ± 0.03 & 0.44 ± 0.02 & 0.54 ± 0.17 & - & 0.48 ± 0.25 & 0.67 ± 0.01 & 0.43 ± 0.21 & 0.64 ± 0.01 & 0.06 ± 0.0 & 0.58 ± 0.0 & 0.4 ± 0.0 & 0.84 ± 0.0 & 0.45 ± 0.06 & 0.41 ± 0.07 & 0.87 ± 0.0 & 0.24 ± 0.01 & 0.63 ± 0.01 \\
 & PRIMO (TTT) & 0.7 ± 0.02 & 0.42 ± 0.01 & 0.66 ± 0.01 & - & 0.6 ± 0.01 & 0.65 ± 0.02 & 0.47 ± 0.14 & 0.56 ± 0.06 & 0.18 ± 0.01 & 0.53 ± 0.01 & 0.38 ± 0.0 & 0.81 ± 0.02 & 0.44 ± 0.06 & 0.44 ± 0.01 & 0.85 ± 0.02 & 0.27 ± 0.01 & 0.59 ± 0.04 \\
\midrule
\multirow[t]{5}{*}{4} & GP& 0.42 ± 0.25 & 0.12 ± 0.17 & 0.21 ± 0.13 & - & 0.16 ± 0.11 & 0.17 ± 0.15 & 0.09 ± 0.18 & 0.14 ± 0.31 & -0.1 ± 0.05 & 0.27 ± 0.07 & 0.14 ± 0.25 & 0.61 ± 0.25 & -0.01 ± 0.13 & 0.04 ± 0.09 & 0.7 ± 0.04 & 0.02 ± 0.06 & 0.31 ± 0.08 \\
 & PRIMO (ICL) & 0.77 ± 0.02 & 0.45 ± 0.0 & 0.68 ± 0.0 & - & 0.62 ± 0.0 & 0.68 ± 0.0 & 0.59 ± 0.0 & 0.63 ± 0.02 & 0.06 ± 0.0 & 0.57 ± 0.0 & 0.41 ± 0.0 & 0.85 ± 0.0 & 0.5 ± 0.0 & 0.45 ± 0.0 & 0.88 ± 0.0 & 0.24 ± 0.0 & 0.65 ± 0.0 \\
\bottomrule
\end{tabular}
\end{adjustbox}
\end{table}
\end{landscape}

\begin{landscape}
\begin{table}[]
    \centering
    \caption{Substitution per-assay results on the hold out set (2/2).}
    \label{tab:sub_results_2}
    \vskip 0.15in
\begin{adjustbox}{max width=1.6\textwidth}

\begin{tabular}{lllllllllllllllllll}

 &  & \rotatebox{270}{GFP\_AEQVI\_Sarkisyan\_2016} & \rotatebox{270}{GRB2\_HUMAN\_Faure\_2021} & \rotatebox{270}{MAFG\_MOUSE\_Tsuboyama\_2023\_1K1V} & \rotatebox{270}{NKX31\_HUMAN\_Tsuboyama\_2023\_2L9R} & \rotatebox{270}{PKN1\_HUMAN\_Tsuboyama\_2023\_1URF} & \rotatebox{270}{Q8WTC7\_9CNID\_Somermeyer\_2022} & \rotatebox{270}{RCD1\_ARATH\_Tsuboyama\_2023\_5OAO} & \rotatebox{270}{RL40A\_YEAST\_Mavor\_2016} & \rotatebox{270}{RL40A\_YEAST\_Roscoe\_2013} & \rotatebox{270}{RL40A\_YEAST\_Roscoe\_2014} & \rotatebox{270}{SCIN\_STAAR\_Tsuboyama\_2023\_2QFF} & \rotatebox{270}{SPA\_STAAU\_Tsuboyama\_2023\_1LP1} & \rotatebox{270}{SQSTM\_MOUSE\_Tsuboyama\_2023\_2RRU} & \rotatebox{270}{SR43C\_ARATH\_Tsuboyama\_2023\_2N88} & \rotatebox{270}{VKOR1\_HUMAN\_Chiasson\_2020\_abundance} & \rotatebox{270}{VKOR1\_HUMAN\_Chiasson\_2020\_activity} & \rotatebox{270}{VRPI\_BPT7\_Tsuboyama\_2023\_2WNM} \\
 & Shots &  &  &  &  &  &  &  &  &  &  &  &  &  &  &  &  &  \\
 \midrule
 \multirow[t]{2}{*}{1} & PRIMO (ICL) & 0.08 ± 0.0 & 0.6 ± 0.0 & 0.17 ± 0.04 & 0.74 ± 0.0 & 0.65 ± 0.01 & 0.05 ± 0.0 & 0.72 ± 0.0 & 0.4 ± 0.0 & 0.45 ± 0.0 & 0.54 ± 0.0 & 0.51 ± 0.01 & 0.2 ± 0.0 & 0.76 ± 0.01 & 0.78 ± 0.05 & 0.45 ± 0.0 & 0.35 ± 0.01 & 0.75 ± 0.09 \\
 & PRIMO (TTT) & 0.08 ± 0.05 & 0.59 ± 0.01 & 0.34 ± 0.08 & 0.64 ± 0.12 & 0.43 ± 0.02 & 0.08 ± 0.01 & 0.64 ± 0.04 & 0.38 ± 0.01 & 0.45 ± 0.02 & 0.47 ± 0.03 & 0.57 ± 0.02 & 0.26 ± 0.02 & 0.72 ± 0.03 & 0.8 ± 0.02 & 0.42 ± 0.01 & 0.34 ± 0.01 & 0.73 ± 0.02 \\
\midrule
\multirow[t]{5}{*}{4} & GP& 0.08 ± 0.22 & 0.23 ± 0.12 & 0.25 ± 0.29 & 0.39 ± 0.34 & 0.25 ± 0.06 & 0.06 ± 0.02 & 0.5 ± 0.07 & 0.11 ± 0.13 & 0.19 ± 0.19 & 0.13 ± 0.14 & 0.29 ± 0.15 & 0.34 ± 0.09 & 0.57 ± 0.05 & 0.64 ± 0.15 & 0.23 ± 0.16 & 0.06 ± 0.1 & 0.5 ± 0.26 \\
\bottomrule
\end{tabular}
\end{adjustbox}
\end{table}
\end{landscape}

\begin{landscape}
\begin{table}[]
    \centering
    \caption{Indel per-assay results on the hold out set (1/2).}
    \label{tab:indel_results_1}
    \vskip 0.15in
\begin{adjustbox}{max width=1.3\textwidth}

\begin{tabular}{lllllllllllllllllll}
 &  & \rotatebox{270}{AMFR\_HUMAN\_Tsuboyama\_2023\_4G3O} & \rotatebox{270}{BLAT\_ECOLX\_Deng\_2012} & \rotatebox{270}{BLAT\_ECOLX\_Firnberg\_2014} & \rotatebox{270}{BLAT\_ECOLX\_Gonzalez\_2019} & \rotatebox{270}{BLAT\_ECOLX\_Jacquier\_2013} & \rotatebox{270}{BLAT\_ECOLX\_Stiffler\_2015} & \rotatebox{270}{CASP3\_HUMAN\_Roychowdhury\_2020} & \rotatebox{270}{CSN4\_MOUSE\_Tsuboyama\_2023\_1UFM} & \rotatebox{270}{D7PM05\_CLYGR\_Somermeyer\_2022} & \rotatebox{270}{DLG4\_HUMAN\_Faure\_2021} & \rotatebox{270}{DLG4\_RAT\_McLaughlin\_2012} & \rotatebox{270}{DNJA1\_HUMAN\_Tsuboyama\_2023\_2LO1} & \rotatebox{270}{DYR\_ECOLI\_Nguyen\_2023} & \rotatebox{270}{DYR\_ECOLI\_Thompson\_2019} & \rotatebox{270}{EPHB2\_HUMAN\_Tsuboyama\_2023\_1F0M} & \rotatebox{270}{ESTA\_BACSU\_Nutschel\_2020} & \rotatebox{270}{FECA\_ECOLI\_Tsuboyama\_2023\_2D1U} \\
 Shots & Method  &  &  &  &  &  &  &  &  &  &  &  &  &  &  &  &  &  \\
\midrule
\multirow[t]{2}{*}{1} & PRIMO (ICL) & 0.69 ± 0.02 & - & - & 0.56 ± 0.03 & - & - & - & 0.81 ± 0.0 & - & - & - & 0.76 ± 0.03 & - & - & 0.7 ± 0.02 & - & 0.47 ± 0.01 \\
 & PRIMO (TTT) & 0.7 ± 0.02 & - & - & 0.57 ± 0.02 & - & - & - & 0.82 ± 0.03 & - & - & - & 0.75 ± 0.03 & - & - & 0.68 ± 0.06 & - & 0.46 ± 0.02 \\
\midrule
\multirow[t]{5}{*}{4} & GP & -0.45 ± 0.21 & - & - & 0.13 ± 0.51 & - & - & - & 0.65 ± 0.05 & - & - & - & 0.65 ± 0.2 & - & - & 0.49 ± 0.1 & - & 0.41 ± 0.04 \\
 & PRIMO (ICL) & 0.71 ± 0.01 & - & - & 0.61 ± 0.0 & - & - & - & 0.8 ± 0.0 & - & - & - & 0.78 ± 0.0 & - & - & 0.69 ± 0.0 & - & 0.47 ± 0.01 \\
 & PRIMO (TTT) & 0.45 ± 0.4 & - & - & 0.62 ± 0.06 & - & - & - & 0.81 ± 0.04 & - & - & - & 0.78 ± 0.01 & - & - & 0.76 ± 0.03 & - & 0.4 ± 0.11 \\
 & RF & 0.24 ± 0.57 & - & - & 0.34 ± 0.32 & - & - & - & 0.4 ± 0.53 & - & - & - & 0.53 ± 0.37 & - & - & 0.48 ± 0.1 & - & -0.02 ± 0.22 \\
\bottomrule
\end{tabular}
\end{adjustbox}
\end{table}
\end{landscape}

\begin{landscape}
\begin{table}[]
    \centering
    \caption{Indel per-assay results on the hold out set (2/2).}
    \label{tab:indel_results_2}
    \vskip 0.15in
\begin{adjustbox}{max width=1.3\textwidth}

\begin{tabular}{lllllllllllllllllll}

 &  & \rotatebox{270}{GFP\_AEQVI\_Sarkisyan\_2016} & \rotatebox{270}{GRB2\_HUMAN\_Faure\_2021} & \rotatebox{270}{MAFG\_MOUSE\_Tsuboyama\_2023\_1K1V} & \rotatebox{270}{NKX31\_HUMAN\_Tsuboyama\_2023\_2L9R} & \rotatebox{270}{PKN1\_HUMAN\_Tsuboyama\_2023\_1URF} & \rotatebox{270}{Q8WTC7\_9CNID\_Somermeyer\_2022} & \rotatebox{270}{RCD1\_ARATH\_Tsuboyama\_2023\_5OAO} & \rotatebox{270}{RL40A\_YEAST\_Mavor\_2016} & \rotatebox{270}{RL40A\_YEAST\_Roscoe\_2013} & \rotatebox{270}{RL40A\_YEAST\_Roscoe\_2014} & \rotatebox{270}{SCIN\_STAAR\_Tsuboyama\_2023\_2QFF} & \rotatebox{270}{SPA\_STAAU\_Tsuboyama\_2023\_1LP1} & \rotatebox{270}{SQSTM\_MOUSE\_Tsuboyama\_2023\_2RRU} & \rotatebox{270}{SR43C\_ARATH\_Tsuboyama\_2023\_2N88} & \rotatebox{270}{VKOR1\_HUMAN\_Chiasson\_2020\_abundance} & \rotatebox{270}{VKOR1\_HUMAN\_Chiasson\_2020\_activity} & \rotatebox{270}{VRPI\_BPT7\_Tsuboyama\_2023\_2WNM} \\
 & Shots &  &  &  &  &  &  &  &  &  &  &  &  &  &  &  &  &  \\
\midrule
\multirow[t]{2}{*}{1} & PRIMO (ICL) & - & - & 0.11 ± 0.03 & 0.75 ± 0.01 & 0.84 ± 0.01 & - & 0.52 ± 0.01 & - & - & - & - & - & 0.61 ± 0.02 & 0.66 ± 0.02 & - & - & 0.85 ± 0.02 \\
 & PRIMO (TTT) & - & - & 0.01 ± 0.05 & 0.69 ± 0.01 & 0.85 ± 0.01 & - & 0.71 ± 0.03 & - & - & - & - & - & 0.61 ± 0.03 & 0.72 ± 0.02 & - & - & 0.84 ± 0.02 \\
\midrule
\multirow[t]{5}{*}{4} & GP & - & - & 0.18 ± 0.35 & 0.39 ± 0.57 & 0.79 ± 0.05 & - & -0.29 ± 0.35 & - & - & - & - & - & 0.32 ± 0.1 & 0.48 ± 0.27 & - & - & 0.05 ± 0.53 \\
 & PRIMO (ICL) & - & - & 0.12 ± 0.01 & 0.76 ± 0.01 & 0.83 ± 0.01 & - & 0.53 ± 0.03 & - & - & - & - & - & 0.61 ± 0.08 & 0.66 ± 0.02 & - & - & 0.87 ± 0.0 \\
 & PRIMO (TTT) & - & - & 0.23 ± 0.12 & 0.76 ± 0.07 & 0.78 ± 0.09 & - & 0.69 ± 0.04 & - & - & - & - & - & 0.59 ± 0.05 & 0.74 ± 0.07 & - & - & 0.83 ± 0.05 \\
 & RF & - & - & 0.09 ± 0.36 & 0.26 ± 0.56 & 0.6 ± 0.19 & - & 0.63 ± 0.04 & - & - & - & - & - & 0.26 ± 0.05 & 0.53 ± 0.11 & - & - & 0.75 ± 0.07 \\
\bottomrule
\end{tabular}
\end{adjustbox}
\end{table}
\end{landscape}



\subsection{Performance by MSA Depth}

\begin{table}[ht]
\centering
\caption{Performance grouped by different levels of MSA depth as defined in ProteinGym.}
\label{tab:msa_depth}
\begin{adjustbox}{max width=1\textwidth}
\begin{tabular}{llccccccc}
\toprule
Depth & Model &  0 &  4 & 8 &  16 &  32 &  64 &  128 \\
\midrule
\multirow[t]{6}{*}{High}  & GP & 0.51 & 0.3 ± 0.03 & 0.38 ± 0.04 & 0.44 ± 0.03 & 0.49 ± 0.01 & 0.54 ± 0.01 & 0.59 ± 0.01 \\
&Ridge & 0.51 & 0.3 ± 0.02 & 0.38 ± 0.04 & 0.45 ± 0.02 & 0.52 ± 0.02 & 0.59 ± 0.01 & 0.65 ± 0.01 \\
&RF & - & 0.26 ± 0.03 & 0.36 ± 0.03 & 0.43 ± 0.02 & 0.48 ± 0.01 & 0.55 ± 0.01 & 0.62 ± 0.01 \\
&PRIMO (ICL) & 0.6 ± 0.03 & 0.64 ± 0.0 & 0.64 ± 0.0 & 0.64 ± 0.0 & 0.64 ± 0.0 & 0.64 ± 0.0 & 0.64 ± 0.0 \\
& PRIMO (TTT) & 0.6 ± 0.01 & 0.57 ± 0.02 & 0.59 ± 0.02 & 0.62 ± 0.01 & 0.65 ± 0.01 & 0.68 ± 0.01 & 0.71 ± 0.01 \\
\midrule
\multirow[t]{6}{*}{Medium} & GP & 0.37 & 0.26 ± 0.03 & 0.34 ± 0.03 & 0.41 ± 0.03 & 0.48 ± 0.02 & 0.56 ± 0.01 & 0.62 ± 0.01 \\
&Ridge & 0.37 & 0.27 ± 0.04 & 0.34 ± 0.03 & 0.43 ± 0.02 & 0.51 ± 0.02 & 0.61 ± 0.01 & 0.69 ± 0.01 \\
&RF & - & 0.24 ± 0.02 & 0.33 ± 0.03 & 0.41 ± 0.01 & 0.48 ± 0.02 & 0.56 ± 0.01 & 0.63 ± 0.0 \\
&PRIMO (ICL) & 0.52 ± 0.01 & 0.53 ± 0.0 & 0.53 ± 0.0 & 0.53 ± 0.0 & 0.53 ± 0.0 & 0.53 ± 0.0 & 0.53 ± 0.0 \\
&PRIMO (TTT) & 0.51 ± 0.01 & 0.48 ± 0.02 & 0.5 ± 0.03 & 0.54 ± 0.02 & 0.6 ± 0.03 & 0.65 ± 0.01 & 0.71 ± 0.0 \\
\midrule
\multirow[t]{6}{*}{Low} & GP & 0.26 & 0.09 ± 0.05 & 0.11 ± 0.05 & 0.18 ± 0.0 & 0.22 ± 0.01 & 0.24 ± 0.01 & 0.29 ± 0.01 \\
&Ridge & 0.26 & 0.12 ± 0.01 & 0.14 ± 0.04 & 0.2 ± 0.03 & 0.26 ± 0.03 & 0.31 ± 0.01 & 0.37 ± 0.01 \\
&RF & - & 0.12 ± 0.04 & 0.16 ± 0.03 & 0.21 ± 0.02 & 0.28 ± 0.03 & 0.32 ± 0.01 & 0.39 ± 0.0 \\
&PRIMO (ICL) & 0.24 ± 0.0 & 0.23 ± 0.0 & 0.23 ± 0.0 & 0.23 ± 0.0 & 0.23 ± 0.0 & 0.23 ± 0.0 & 0.23 ± 0.0 \\
&PRIMO (TTT) & 0.25 ± 0.01 & 0.26 ± 0.03 & 0.23 ± 0.03 & 0.28 ± 0.03 & 0.33 ± 0.03 & 0.39 ± 0.01 & 0.45 ± 0.01 \\

\bottomrule
\end{tabular}
\end{adjustbox}
\end{table}

\section*{Software and Data}

Code is available at \url{https://anonymous.4open.science/r/PRIMO-D3F9/README.md}

\end{document}